\documentclass[english,reprint,prl]{revtex4-1}
\usepackage[T1]{fontenc}
\usepackage[latin9]{inputenc}
\usepackage{amsmath}
\usepackage{amsthm}
\usepackage{amssymb}
\usepackage{stmaryrd}
\usepackage{mathtools}
\usepackage{graphicx}
\usepackage[hidelinks]{hyperref}
\usepackage{xargs}[2008/03/08]
\usepackage[capitalize]{cleveref}

\makeatletter
\theoremstyle{plain}
\newtheorem{thm}{\protect\theoremname}

\newtheorem{prop}{Proposition}

\DeclareMathOperator*{\argmin}{\arg\,\min}
\DeclareMathOperator*{\argmax}{\arg\,\max}

\RequirePackage[dvipsnames,usenames]{xcolor}

\crefname{appsec}{Appendix}{Appendices}
\crefname{appsubsec}{Appendix}{Appendices}

\usepackage{newtxtext}

\newcommand\allDistSet{\mathcal{P}}
\newcommand\distSet{\Omega}

\crefname{thm}{Theorem}{Theorems}
\crefname{lem}{Lemma}{Lemmas}
\crefname{prop}{Proposition}{Propositions}
\crefname{defn}{Definition}{Definitions}
\crefname{rem}{Remark}{Remarks}

\makeatother

\usepackage{babel}
\providecommand{\lemmaname}{Lemma}
\providecommand{\theoremname}{Theorem}

\begin{document}
\global\long\def\ddt{{\textstyle \frac{d}{dt}}}%
\global\long\def\mult{\eta}%
\global\long\def\ft{\tau}%

\global\long\def\DKL{D}%

\newcommand\R{W}

	\newcommand\Enot{E^\varnothing}
	\newcommandx\EnotX[1][usedefault, addprefix=\global, 1=\ii]{E^\varnothing_{#1}}%

	\newcommandx\Rt[1][usedefault, addprefix=\global, 1=t]{\R({#1})}%
	\newcommandx\Et[1][usedefault, addprefix=\global, 1=t]{E({#1})}%
	\newcommandx\pxt[2][usedefault, addprefix=\global, 1=\ii, 2=t]{ p_{#1}({#2})}%
	\newcommandx\trans[3][usedefault, addprefix=\global, 1=\R]{{#1}_{#3 #2}(t)}%
	\newcommandx\pt[1][usedefault, addprefix=\global, 1=t]{p({#1})}%
	\newcommandx\transF[3][usedefault, addprefix=\global, 1=\R]{{#1}_{#3 #2}}%
	\newcommandx\px[1][usedefault, addprefix=\global, 1=\ii]{p_{#1}}%
	\newcommandx\Ex[1][usedefault, addprefix=\global, 1=\ii]{E_{#1}}%
	\newcommandx\Ext[1][usedefault, addprefix=\global, 1=\ii]{E_{#1}(t)}%
	\global\long\def\ii{i}
	\global\long\def\jj{j}
	\global\long\def\Eit{E_{\ii}(t)}%
	\global\long\def\Ejt{E_{\jj}(t)}%

\global\long\def\EP{\Sigma}%
\global\long\def\EPr{\dot{\EP}}%
\global\long\def\prot{\Gamma}%
\global\long\def\protSet{\{(\Et,\Rt):t\in[0,\ft]\}}
\global\long\def\protSetEE{\{(\Et,\Rt):t\in[0,\ft],\Et\in\EE\}}
\global\long\def\protSetEERR{\{(\Et,\Rt):t\in[0,\ft],\Et\in\EE,\Rt\in\RR\}}
\global\long\def\protAlpha{\Gamma_{\alpha}}%
\global\long\def\EE{\mathcal{E}}%

\global\long\def\pf{p'}%
\global\long\def\ps{p}%
\newcommandx\pxs[1][usedefault, addprefix=\global, 1=\ii]{ \ps_{#1} }
\newcommandx\pxf[1][usedefault, addprefix=\global, 1=\ii]{ \pf_{#1} }

\global\long\def\pptp{\pf\to\ps}%
\global\long\def\ptpp{\ps\to\pf}%

\global\long\def\Wji{\trans{\ii}{\jj}}%
\global\long\def\Wij{\trans{\jj}{\ii}}%
\global\long\def\WjiF{\transF{\ii}{\jj}}%
\global\long\def\WijF{\transF{\jj}{\ii}}%

\global\long\def\odd{\mathrm{o}}%
\global\long\def\even{\mathrm{e}}%
\global\long\def\scO{\psi^{\odd}}%
\global\long\def\scE{\psi^{\even}}%
\global\long\def\scOmax{\phi^{\odd}}%
\global\long\def\scEmax{\phi^{\even}}%
\global\long\def\klprojBase{\Pi}%

\global\long\def\PathsIJ{\mathcal{P}(\ii,\jj)}%

\newcommand{\pmin}{p_{\min}}
\newcommandx\maxCap[1][usedefault, addprefix=\global, 1=\EE]{\mathsf{C}(#1)}%
\newcommand{\pathcap}{\mathsf{c}}

\global\long\def\piE{\pi^{E}}%

\global\long\def\RR{\mathcal{\R}}%

\newcommandx\klproj[1][usedefault, addprefix=\global, 1=p]{\Pi(#1)}%
\newcommandx\klprojT[1][usedefault, addprefix=\global, 1=t]{\klproj[\pt[#1]]}%

\newcommandx\klprojPS{\klproj[\ps]}%
\newcommandx\klprojPF{\klproj[\pf]}%

\newcommand\bndconst{\gamma(\EE)}

\title{Entropy production given constraints on the energy functions}
\author{Artemy Kolchinsky}
\altaffiliation{artemyk@gmail.com}
\affiliation{Santa Fe Institute, Santa Fe, New Mexico}

\author{David H. Wolpert}
\altaffiliation{
Complexity Science Hub, Vienna; Arizona State University, Tempe, Arizona; 
\texttt{http://davidwolpert.weebly.com}}
\affiliation{Santa Fe Institute, Santa Fe, New Mexico}

\begin{abstract}
We consider the problem of driving a finite-state classical system from some initial distribution $\ps$ to some final distribution $\pf$ with vanishing entropy production (EP), under the constraint that the driving protocols can only use some limited set of energy functions $\EE$. Assuming no other constraints on the driving protocol, we derive a simple condition that guarantees that such a transformation can be carried out, which is stated in terms of the smallest probabilities in $\{\ps,\pf\}$ and a graph-theoretic property defined in terms of $\EE$. Our results imply that a surprisingly small amount of control over the energy function is sufficient (in particular, any transformation $\ptpp$ can be carried out as soon as one can control some one-dimensional  parameter of the energy function, e.g., the energy of a single state). We also derive a lower bound on the EP under more general constraints on the transition rates, which is formulated in terms
of a convex optimization problem. 
\end{abstract}
\maketitle

\section{Introduction}

Entropy production (EP) refers to the total increase of the entropy of a system and its environment during a physical process. EP is the fundamental measure of thermodynamic inefficiency~\cite{seifert2012stochastic}, and in particular, the amount of work that can be extracted during a transformation between two thermodynamic states  decreases as EP increases~\cite{esposito2011second}.  For this reason, one of the central issues in thermodynamics involves characterizing the minimal amount of EP needed to transform a system between two thermodynamic states.

For concreteness, consider a system coupled to a work reservoir and a single heat bath at inverse temperature $\beta$. %
Suppose that one wishes to drive this system from some initial distribution $\ps$ at time $t=0$ to some final distribution $\pf$ at time $t=\ft$ while minimizing EP. In general, this can be accomplished by driving the 
system along a 
continuous trajectory of energy functions that sends %
\begin{align}
\Et[0]=-\beta^{-1} \ln \ps \;\to \; \Et[\ft]=-\beta^{-1} \ln \pf.
\label{eq:curve0}
\end{align}
In the quasistatic limit $\ft\to\infty$, this will carry out the transformation $\ptpp$ up to an arbitrary accuracy and for an arbitrarily small amount of EP~\cite{takara_generalization_2010,parrondo2015thermodynamics}.
However, implementing such a driving protocol may be impossible in many real-world situations, where there are often strong %
limitations on the ability to control the energy function.

Here we investigate which transformations can be carried out without EP, assuming that one is constrained to
use some limited set of energy functions, but is otherwise free to use any thermodynamically consistent driving protocol.   
We consider a finite-state classical 
system which %
evolves according to a Markovian master
equation over time $t\in[0,\ft]$,
\begin{align}
\ddt \pxt=\sum_{\jj (\ne \ii)}\big[\Wij\pxt[\jj]-\Wji\pxt\big],
\label{eq:me}
\end{align}
where $\pxt$ is the probability of state $\ii$ at time $t$ and $\Wji$ is the transition rate from state $\ii$ to state $\jj$ at time $t$. 
As standard in stochastic thermodynamics~\cite{van2013stochastic}, we assume that %
the rate matrix obeys local detailed balance (LDB) at all times $t$, so that the transition rates can be written in the following form~\cite[Ch.~2]{maes2018non}:
\begin{align}
\Wji=\psi_{ji}(t) e^{\beta[\Ext[\ii]-\Ext[\jj]]/2},
\label{eq:paramRates}
\end{align}
where $\Et$ is the energy function at time $t$ and $\psi_{ji}(t)=\psi_{ij}(t)\ge 0$ is a symmetric positive function that controls the overall timescale of transitions between states $\ii$ and $\jj$,  sometimes called the \emph{activity}~\cite{maes2018non}.   \cref{eq:paramRates} implies that the rate matrix $W(t)$ has a Boltzmann equilibrium distribution, $\pi^{\Et} = e^{-\beta\Et}/Z$. 
We use the term \emph{(driving) protocol} to refer to a time-dependent
trajectory of LDB-obeying rate matrices and energy functions, $\prot = \protSet$. (The inverse temperature $\beta$ associated with the driving protocol is left  implicit in our notation).

Suppose that there is some limited set of  energy functions $\EE$ that one can impose on the system. 
Given some desired transformation $\ptpp$, we ask whether there is a driving protocol that carries out this transformation for a vanishing amount of EP, while obeying the constraint that $\Et\in\EE$ at all $t\in[0,\ft]$. %

Our first result, presented  in \cref{sec:first}, is a simple sufficient condition %
that guarantees that 
there is such a protocol. 
This sufficient condition is stated in terms of the minimum probability values in $\ps$ and $\pf$, as well as a simple graph-theoretic property defined in terms of $\EE$. %
As we show, this result implies that a surprisingly limited amount of control over the energy function is sufficient to carry out an arbitrary transformation $\ptpp$ with vanishing EP. 
In particular, it typically suffices to manipulate the energy along some arbitrary  one-dimensional control parameter (for example, the magnetic field applied to an Ising model, or the energy assigned to some arbitrary single state). 
This can be contrasted to the driving protocol in \cref{eq:curve0}, which requires access to energy functions that are tailored
to both $\ps$ and $\pf$, rather than being restricted to an arbitrary fixed one-dimensional set. %

As a motivating example, consider a system
composed of two subsystems $X \times Y$, and suppose that
$\EE$ only contains decoupled energy functions such as
\begin{align}
\Ex[x,y]=\Ex[x]+\Ex[y].\label{eq:additiveE}
\end{align}
Imagine that one wants to bring the system from some initial equilibrium distribution $\ps=\piE$ for some $E\in\EE$ to some final distribution $\pf$. Suppose that the final distribution has correlations between the two subsystems, such that the associated
mutual information obeys $I_{\pf}(X;Y)>0$.  For any decoupled energy function as in \cref{eq:additiveE}, the equilibrium distribution is product distribution, with zero mutual
information. Intuitively, it might seem impossible to introduce correlations between 
$X$ and $Y$, as one must
do to bring the system to the desired ending distribution $\pf$, while driving the system using only decoupled energy functions.  
As we show below, this intuition is wrong: there is a driving protocol that increases correlations between $X$ and $Y$, incurs vanishing EP, and only uses decoupled energy functions (in fact, it suffices to only manipulate the energy of a single state of a single subsystem).

Conversely, imagine that one wants to use the same set of decoupled energy functions
$\EE$ to bring the system from some correlated initial distribution $\ps$, such that $I_{\ps}(X;Y)>0$, to a final equilibrium distribution  
$\pf = \piE$ for some $E\in\EE$. %
One way to achieve this transformation would be to apply the energy function $E$ to the system and then let it relax freely  to the final equilibrium distribution $\piE$. However,  under such a free relaxation, any mutual information in the initial distribution would be dissipated as EP~\footnote{It is known that for a free relaxation from an initial distribution \unexpanded{$\ps$} to a final equilibrium distribution  \unexpanded{$\pf=\piE$}, the EP obeys \unexpanded{$\EP=\DKL(\ps \Vert \piE )$}~\cite{schloglStatisticalFoundationThermodynamic1967}. In addition, we can rewrite \unexpanded{$\DKL(\ps\Vert \piE) = I_{\ps}(X;Y) + \DKL(p_X p_Y \Vert \piE_X \piE_Y) \ge I_{\ps}(X;Y)$}. %
Combining implies \unexpanded{$\EP\ge I_{\ps}(X;Y)$}.}. %
Intuition might suggest that for any protocol that uses decoupled energy functions, EP can always be bounded  
in terms of the mutual information $I_{\ps}(X;Y)$. 
(In fact, such bounds have been previously derived under related but different assumptions, involving constraints not only on the energy functions but also some other parameters that determine the rate matrices~\cite{wolpert2020uncertainty,wolpert2020minimal,wolpert2019stochastic,wolpert2020thermodynamic,boyd2018thermodynamics,kolchinsky2020entropy}.) It turns out, however, that initial correlations do not have to be dissipated as EP, even if only decoupled energy functions are available. Rather, as we show below, there is a driving protocol that carries out 
the transformation $\ps\to \piE$ without EP while only using decoupled energy functions.

It is important to emphasize that our first result is proved by construction, in which we assume that there are no other constraints on the protocol beyond that the energy function must belong to $\EE$. In particular, we assume that one can impose a separation of timescales and fix any subset of the transition rates $\Wij$ to zero (i.e., by setting the corresponding activity parameters to zero). %
For this reason, our construction can be seen as an illustration of the limits of what is allowed by the laws of stochastic thermodynamics given constraints on the energy functions (rather than an example of a protocol that is easy to implement in practice).

Of course, in many real-world scenarios, there are also other constraints on the rate matrices, and in particular %
it may not be possible to set certain rates to zero. 
In our second result, presented  in \cref{sec:second},  
we derive a lower bound on the minimal EP that must be incurred when carrying out the transformation $\ptpp$ given more general constraints on the available rate matrices. This bound is stated in terms of a convex optimization problem, which can sometimes be solved using standard numerical techniques.

For simplicity of presentation,  we introduce our results  in the context of a system coupled to a single heat bath and subject to conservative forces, as arising from some time-dependent energy potential $\Et$. %
However, our results can be generalized to systems coupled to multiple reservoirs and/or subject to  non-conservative forces. In \cref{app:nonconservative}, we present a generalization of our first result for non-conservative forces, including a simple sufficient condition that guarantees that a given transformation $\ptpp$ can be carried out for a vanishing amount of EP, assuming that there are constraints on both the energy functions and the non-conservative forces that can be applied to a given system. In \cref{app:thm2multiplebaths}, we present a generalization of our second result, involving a bound on EP under a general set constraints on the rate matrices, for a system coupled to any number of reservoirs.

\section{Prior work}

This paper extends our recent work~\citep{kolchinsky2020entropy}, in which we derived 
bounds on EP for systems that evolve under rate matrices that have symmetrical, modular, or coarse-grained structure.

In other relevant literature, Ref.~\cite{wilming_second_2016} analyzed EP  in the presence of constraints on the Hamiltonian for a quantum system coupled to a finite-sized heat bath. That paper derived bounds for special protocols that consists of sequences of unitary transformations of the system+bath and total relaxations of the system to an equilibrium state. In contrast, we analyze a classical system coupled to an idealized (i.e., infinite) reservoir and consider a much broader set of protocols than just sequences of the two kinds of operation
considered in~\cite{wilming_second_2016}.

The derivation of our first main result, \cref{thm:a}, is based on decomposing a general transformation $\ptpp$ into a sequence of simpler transformation over two-state subsystems. Similar constructions have been used to analyze allowed transformations and work extraction in quantum systems~\cite{lostaglio2018elementary,baumerImperfectThermalizationsAllow2019,perrySufficientSetExperimentally2018}.  In classical systems, similar constructions have been used to study how a logical input-output map can be implemented  using a continuous-time Markovian process~\cite{owen_number_2018,wolpert2019space,korzekwa2021quantum}. 

Finally, at a broader level, our paper complements previous analysis of optimal heat and EP generation under realistic constraints on the driving protocols, such as finite time~\citep{esposito2010finite,sivak2012thermodynamic,shiraishi_speed_2018,gomez2008optimal,then2008computing,zulkowski2014optimal,schmiedl2007optimal,baumerImperfectThermalizationsAllow2019,remlein2021optimality,aurell2011optimal,aurell2012refined}, stochastically fluctuating control~\cite{machta2015dissipation}, finite-sized work reservoirs \citep{verley_work_2014},  and a limited set of ``free operations'' as considered in quantum resource theory of thermodynamics~\cite{gour2015resource}.

\newcommand\ff{\varphi}
\newcommandx\egS[3][usedefault, addprefix=\global, 1=\ii,2=\jj,3=\EE]{{ \transF[g] {#2} {#1}    }(#3)}%

\section{Notation and Preliminaries}
\label{sec:prelims}

We consider a system with $n$ states which undergoes a driving protocol.
We use notation like $p,p(t),\ldots \in\mathbb{R}^n_{+}$ to refer to (possibly time-dependent) probability distributions over states. Similarly, we use notation like $E,E(t),\ldots \in \mathbb{R}^n$ to refer to (possibly time-dependent) energy functions over the states. The probability of a particular state $\ii$ is indicated with subscript notation ($p_i,p_i(t)$, etc.), and similarly for the energy level of a particular state ($E_i,E_i(t)$, etc.). We use $\delta_{i,j}$ to indicate the Kronecker delta function.
 
We use the Kullback-Leibler (KL) divergence, an information-theoretic non-negative measure of the difference between distributions. The KL divergence from distribution $p$ to  distribution $q$ is defined as 
\begin{align}
\DKL(p\Vert q) = \sum_i p_i \ln \frac{p_i}{q_i}.
\label{eq:kldef}
\end{align}

Given a driving protocol $\prot=\protSet$, 
the incurred entropy production (EP) on initial distribution $p$  %
is given by
\begin{align}
\EP(p,\Gamma) = \int_0^\ft \EPr(p(t),W(t)) \,dt,
\label{eq:EPdefInt}
\end{align}
where $p(t)$ is the distribution at time $t$ (as  determined by the master equation in \cref{eq:me} under the initial condition $p(0)=p$) and $\EPr(p(t),W(t))$ is the instantaneous EP rate at time $t$.  In stochastic thermodynamics, the instantaneous EP rate incurred by some rate matrix $W$ and distribution $p$ is given by~\cite{schnakenberg1976network,esposito2010three}
\begin{align} 
\EPr(p,W) &= \frac{1}{2}\sum_{\ii, \jj}{ (\px\WjiF - \px[\jj]\WijF)}  \ln \frac{\px\WjiF}{\px[\jj]\WijF}\ge 0.
\label{eq:eprate0}
\end{align}

\section{EP under energy constraints}
\label{sec:first}

Consider some driving protocol $\prot=\protSet$ which obeys LDB, and suppose that only a %
restricted set of energy functions is available, so that $\Et \in \EE\subseteq \mathbb{R}^n$ at all times $t\in[0,\ft]$. 
Given this constraint, we investigate whether a given transformation $\ptpp$ can be carried out while achieving  an arbitrarily small amount of EP.
We make the weak assumption that the set $\EE$ is closed and \emph{path-connected}, meaning that any two elements of $\EE$ can be connected by a continuous curve in $\EE$.

Before proceeding, we define the concept of the \emph{controllable energy gap}, which will play a central role in our analysis. The controllable energy gap between a pair of states $\ii,\jj$ is
\begin{align}
\egS = \min\Big\{ \max_{E\in\EE}\, (\Ex - \Ex[\jj]), \max_{E\in\EE} \,(\Ex[\jj] - \Ex)\Big\}.
\label{eq:egSdef}
\end{align}
In words, $\egS$ quantifies how much the energy gap between states $\ii$ and $\jj$ can be varied by choosing among different $E\in\EE$. As an illustration, if the set of available energy functions is a line segment, $\EE=\{\lambda \Enot : \lambda \in [-1,1]\}$ for some fixed energy function $\Enot$, then $\egS = |\EnotX[\jj]-\EnotX[\ii]|$. 

\begin{figure*}
\begin{centering}
\includegraphics[width=1.8\columnwidth]{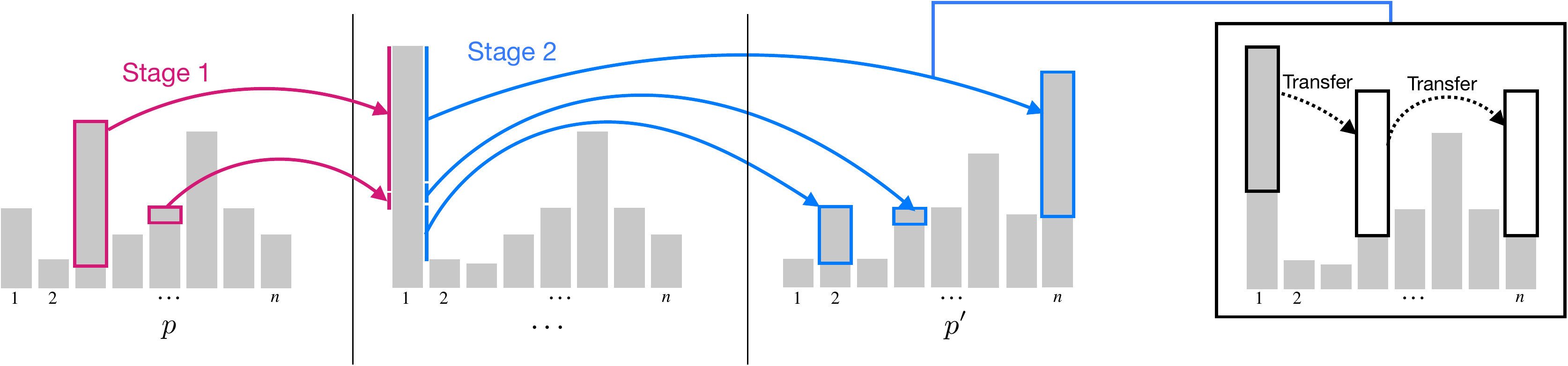}
\par\end{centering}
\caption{A two stage construction used to carry out a transformation $\ptpp$ using a sequence of transfers. In  stage 1, for all states $\ii$ with excess probability, $\pxs > \pxf$,  $\pxs - \pxf$ of probability is moved into a buffer state (state 1). In stage 2, $\pxf - \pxs$ of probability from the buffer state is moved to all states $\ii$ with $\pxs < \pxf$.  Each probability move (red and blue arrows) is done using a sequence of state-to-state transfers along a path (box on right).
\label{fig:transfers}}
\end{figure*}

We first consider the special case where going from $\ps$ to $\pf$ only involves moving $\Delta$ probability from some state $a$ to some other state $b$, so that
\begin{align}
\pxf = \pxs +  \Delta (\delta_{b,\ii} - \delta_{a,\ii}).
\label{eq:dyn0}
\end{align}
We then show how to construct a special
driving protocol %
which carries out this transformation while incurring an arbitrarily small amount of EP, and
while only using energy functions $E \in \EE$ along with rate matrices that obey LDB. 

To begin, recall
that one can {always} set the transition rates going both forward and 
backward between any two states $\ii,\jj$ to zero without violating  LDB (i.e., by setting the corresponding activity parameters $\psi_{ij}(t)=\psi_{ji}(t)=0$ in \cref{eq:paramRates}). 
This allows us to construct a driving protocol such that at all $t$, only transitions between states $a$ and $b$ are allowed, because $\Wji = 0$ whenever $\ii \not\in \{a,b\}$ or $\jj \not\in \{a,b\}$. %
In this case, %
 only the pair of states $a,b$ contributes to the EP rate in \cref{eq:eprate0}, allowing us to write %
 the EP rate at time $t$ as %
\begin{multline}
\EPr(p(t),W(t)) = %
{(p_a(t) W_{ba}(t) - p_b(t) W_{ab}(t))} \\\times{\Big[\ln \frac{\pxt[a]}{\pxt[b]} \!-\! \beta(\Ext[a]-\Ext[b])\Big]},
\label{eq:eprate0ldbAB}
\end{multline}
where we used \cref{eq:eprate0} as well as LDB, \cref{eq:paramRates}.
Now suppose that the controllable energy gap between $a$ and $b$  obeys
\begin{align}
\egS[a][b]>  \beta^{-1} \max\left\{ \left|\ln \frac{\pxs[a]}{\pxs[b]}\right|,  \left|\ln \frac{\pxf[a]}{\pxf[b]}\right|\right\}.
\label{eq:cg2}
\end{align}
Given the definition of $\egS[a][b]$ in \cref{eq:egSdef}, as well as the assumption that $\EE$ is path-connected, there must be some $E\in\EE$ such that $\Ex[a]-\Ex[b]$ is equal to any desired value between  $ \beta^{-1}\vert \ln ({\pxs[a]}/{\pxs[b]})\vert$ and $ \beta^{-1}\vert \ln ({\pxf[a]}/{\pxf[b]})\vert$.
Intuitively, this suggests that at any time $t$, one can choose the energy function within $\Et\in\EE$ so that the bracketed term in  \cref{eq:eprate0ldbAB}
becomes arbitrarily small.  In fact, in the quasistatic limit of $\ft\to\infty$, the energy function can be varied in such a way that the total EP incurred over $t\in[0,\ft]$ becomes arbitrarily small. 
By formalizing this intuition, we derive the following result, which is proved in \cref{app:transfer}. 

\begin{prop}
\label{lem:transfer}
If \cref{eq:dyn0,eq:cg2} hold,  
then for any $\epsilon>0$, there is a protocol $\prot=\protSet$
that sends $\ptpp$ while obeying $\EP(p,\prot) \le \epsilon$ and $E(t)\in\EE$ at all $t$.  
\end{prop} 
\noindent
We refer the driving protocol constructed in the proof of \cref{lem:transfer}, which moves probability between two states while incurring a vanishing amount of EP, as a \emph{transfer}.

\newcommand\pathIK[1]{x_{#1}}

We are now ready to prove our first main result, 
which states that one can drive the system from any initial distribution $\ps$ to any  final distribution $\pf$ %
by chaining together an
appropriate sequence of transfers.  Moreover, because each transfer can be carried out quasistatically and thereby incur an arbitrarily small amount of EP, the overall sequence of transfers can also be made to incur an arbitrarily small amount of EP.

There are many protocols that can carry out the transformation $\ptpp$ using a sequence of transfers. %
One relatively simple one involves a two stage process, illustrated in \cref{fig:transfers}. 
In the first stage, we pick one particular ``buffer state'' (without loss of generality, this can be state 1) to 
accumulate probability from states with excess probability (this accumulation  is done via a sequence of
 transfers). In the second stage, the probability accumulated in the buffer state is distributed to states that need probability (this distribution is again done via a sequence of transfers). This procedure is outlined more formally as follows:

\vspace{5pt}
\noindent {Stage 1)} Consider in turn each state $\ii$ such that $\pxs>\pxf$. For each such $\ii$, we will move $\Delta_\ii=\pxs-\pxf$ of probability from state $\ii$ to state $1$ (if $\ii=1$, do nothing). To do so, select some \emph{path} (sequence of states) of length $\ell_\ii$ that starts on state $\ii$ and ends on state 1,
\[
\vec{x}= (\pathIK{1} = \ii \to \pathIK{2} \to \dots \to \pathIK{\ell_\ii -1} \to \pathIK{\ell_\ii} = 1),
\]
and then run $\ell_\ii-1$ transfers, each of which  moves $\Delta_\ii$ of probability from state $\pathIK{k} $ to state $\pathIK{k+1}$ during a temporal interval of length $\ft$.

\vspace{5pt}
\noindent {Stage 2)} Consider in turn each state $\ii$ such that $\pxs<\pxf$. For each such $\ii$, move $\Delta_\ii=\pxf-\pxs$ of probability from state $1$ to state $\ii$ (if $\ii=1$, do nothing). To do so, select some path of length $\ell_\ii$ that starts on state 1 and ends on state $\ii$,
\[
\vec{x}= (\pathIK{1} = 1 \to \pathIK{2} \to \dots \to \pathIK{\ell_\ii -1} \to \pathIK{\ell_\ii} = \ii),
\]
and then run $\ell_\ii-1$ transfers, each of which  moves $\Delta_\ii$ of probability from state $\pathIK{k} $ to state $\pathIK{k+1}$ during a temporal interval of length $\ft$.

\vspace{5pt}

It is easy to see that this construction will transform $\ps$ to $\pf$. In addition, as mentioned above, each individual transfer can be made to have an arbitrarily small amount of EP by taking $\ft$ (the time taken by each transfer) to be sufficiently large. Finally, because the construction involves only a finite number of transfers~\footnote{There are at most \unexpanded{$n-1$} states that have either \unexpanded{$\pxs<\pxf$} or \unexpanded{$\pxs>\pxf$} and are not state 1. Furthermore, each path can have at most \unexpanded{$\ell=n-1$} steps. Thus, the number of transfers involved in the two-stage construction is no greater than \unexpanded{$(n-1)^2$}.}, the overall procedure can be made to incur an arbitrarily small amount of EP.

The construction described above can be carried out as long as the controllable energy gaps are large enough for each  transfer to be feasible --- in other words, as long as the inequality in \cref{eq:cg2} is satisfied for each transfer. %
We now derive a simple sufficient condition that guarantees that this inequality is satisfied for each transfer.

\begin{figure}[b]
\begin{centering}
\includegraphics[width=0.4\columnwidth]{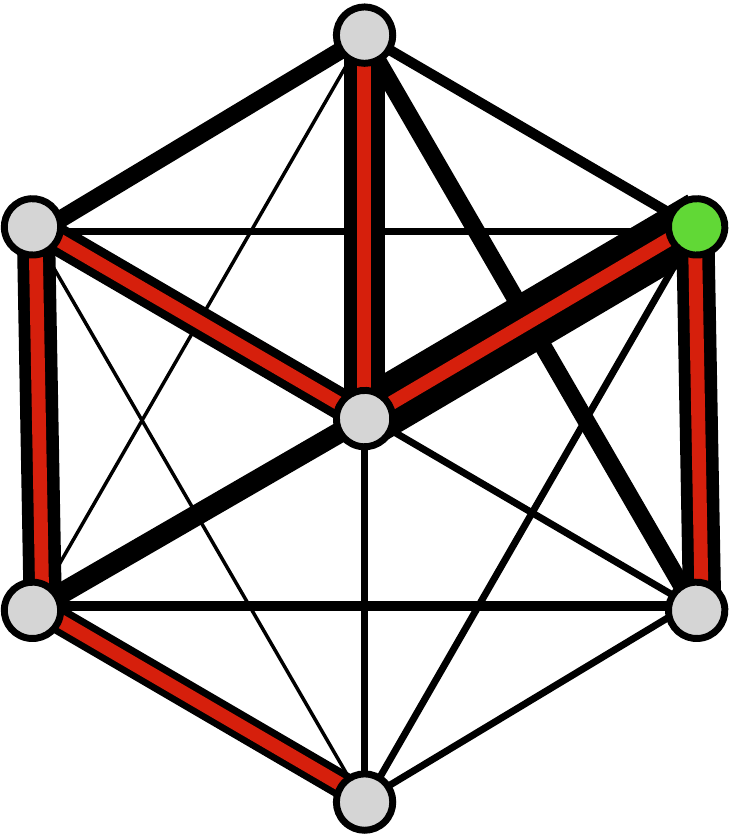}
\par\end{centering}
\caption{Consider a weighted graph with $n$ vertices, where each vertex represents a state of the system and edge weights are given by the controllable energy gaps $\egS$. If the two-stage construction shown in \cref{fig:transfers} chooses paths to/from the buffer state (green node) that lie within the maximum spanning tree of the graph (red edges), then the controllable energy gap across each transfer will be lower-bounded by the maximum capacity of the graph, as stated in \cref{eq:condF}.
\label{fig:mst}}
\end{figure}

First, observe that at no step in the above construction does the probability concentrated in any state $i$ drop below $\min \{\pxs,\pxf\}$. This implies that for any time $t$ at which a transfer begins or ends, the probability concentrated in state $i$ obeys %
\begin{align}
\pxt \ge   \min \{\pxs, \pxf\}\ge \min_{\jj} \min \{\pxs[\jj], \pxf[\jj]\}.
\label{eq:mbound}
\end{align}
Next, observe that in both stages of the above construction, one has the freedom to choose the specific path to/from state 1 and state $\ii$.  In fact, one can choose paths optimally so that the controllable energy gap $\egS[\pathIK{k}][\pathIK{k+1}]$ across each transfer is as large possible, since that increases the set of transformations allowed by \cref{eq:cg2}.   %
Let  $\PathsIJ$ be the set of all paths from state $\ii$  to state $\jj$, and define the  \emph{capacity} of any such path $\vec{x}=(\pathIK{1}=1,\pathIK{2},\dots,\pathIK{m}=j)\in \PathsIJ$ as 
the minimum controllable energy gap in the path,  $ \pathcap(\vec{x}) := \min_k \egS[\pathIK{k}][\pathIK{k+1}]$. 
The capacity of the optimal path between the worst-case pair of states is then given by %
\begin{align}
\maxCap := \min_{\ii \ne \jj} \max_{\vec{x} \in \mathcal{P}(\ii,\jj)} \pathcap(\vec{x}).
\label{eq:maxCapDef}
\end{align}
In graph theory, $\maxCap$ is known as the \emph{maximum capacity} of the undirected graph with $n$ vertices the edge weights of which are specified by $\egS$~\cite{pollack1960letter}.  A classical result states that if paths are chosen from a \emph{maximum spanning tree} of the graph, then no path will have capacity less than $\maxCap$~\cite{hu1961letter}. This means that  in practice %
$\maxCap$ can be calculated as
\begin{align}
\maxCap := \min_{(\ii,\jj)\in T} \egS
\label{eq:mst2}
\end{align}
where $T$ is any maximum spanning tree. %

To summarize, suppose that the two-stage construction described above chooses paths between state $\ii$ and state 1 that lie along a maximum spanning tree of the graph with edge weights $\egS$, as illustrated in \cref{fig:mst}. Then, by \cref{eq:mst2}, %
the 
controllable energy gap involved in any transfer will obey
\begin{align}
\egS[\pathIK{k}][\pathIK{k+1}] \ge \maxCap.
\label{eq:condF}
\end{align}
Finally, suppose that the probabilities in $\pxs$ and $\pxf$ satisfy the following condition: %
\begin{align}
 \min_{\ii} \min \{\pxs, \pxf\} > e^{-\beta \maxCap}.
\label{eq:cond1}
\end{align}
Combining this inequality with \cref{eq:mbound,eq:condF} and rearranging implies $\egS[\pathIK{k}][\pathIK{k+1}] > -\beta^{-1} \ln \pxt[\ii]$ for any state $i$ and any time point $t$ at which a transfer begins or ends. Since $\pxt[\ii] \in[0,1]$ for all $\ii$ and $t$, this in turn implies that  \cref{eq:cg2} 
holds for each transfer. %
This leads to our first main result, which is proved informally using the construction and arguments outlined above.

\begin{thm}
\label{thm:a}%
If \cref{eq:cond1} is satisfied, %
then for any $\epsilon>0$ there is a protocol $\prot=\protSet$
that carries out $\ptpp$ while obeying $\EP(p,\prot)\le \epsilon$ and $E(t)\in\EE$ at all $t$.
\end{thm}

The graph-theoretic quantity $\maxCap$ measures the ability of  the set of available energy functions $\EE$ to implement arbitrary transformations. As expected, this quantity is invariant if any $E \in\EE$  is shifted by a constant (since $\egS$ is invariant under shifts such as $E\to E+\lambda$) and scales multiplicatively with multiplicative scaling of the energy functions ($\maxCap \to \lambda \maxCap$ when $E \to \lambda E$ for all $E \in \EE$).  %

\begin{figure}[b]
\begin{centering}
\includegraphics[width=.4\columnwidth]{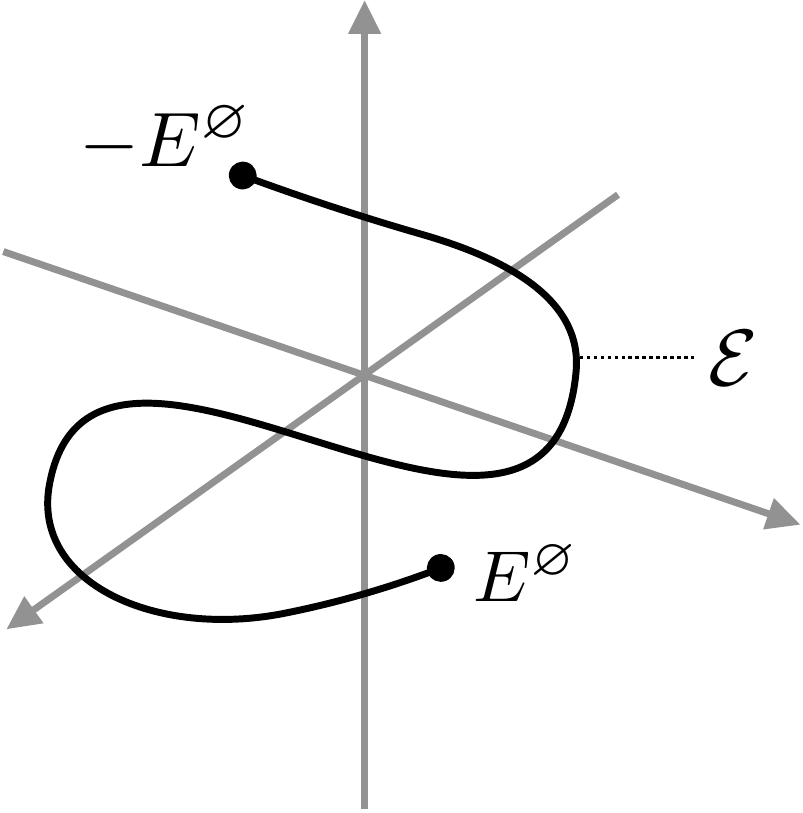}
\par\end{centering}
\caption{
Suppose that the set of available energy functions $\EE \subseteq \mathbb{R}^n$ contains some fixed one-dimensional curve of energy functions from $\Enot$ to $-\Enot$. Then any transformation $\ptpp$ can be carried out for vanishing EP, as long as $\ps$ and $\pf$ have full support and $\Enot$ has a  large enough gap between highest and lowest energy levels, \cref{eq:eucCap}. 
\label{fig:oneD}}
\end{figure}

\cref{thm:a} implies that many transformations can be carried out  with vanishing EP even if only a very restricted set of energy functions is available.  In particular, suppose that  $\EE$ contains a continuous curve that connects some fixed energy function $\Enot$ and its negation $-\Enot$, as shown in \cref{fig:oneD}. 
This curve might involve separately varying the energy level of an individual state, or it might involve varying the entire energy function along a one-dimensional manifold without being able to vary the energy of individual states.  
In \cref{app:eucCap}, we show that for any such $\EE$, 
\begin{align}
\maxCap \!\ge\! \min_{\ii} \max_{\jj} |\EnotX[\ii] - \EnotX[\jj]|\!\ge\! \frac{\max_i \EnotX -\min_i \EnotX}{2},
\label{eq:eucCap}
\end{align}
We also show that the first inequality is tight whenever $\EE$ is a line segment ($\EE = \{ \lambda \Enot  : \lambda \in [-1,1]\}$).  
\cref{eq:eucCap} implies that even a single dimension of control over the energy function suffices to carry out any transformation, as long as $\Enot$ has a sufficiently large range of energy values. In particular, $\maxCap \to \infty$ as the energetic gap between the highest and lowest energy state increases ($\max_i \EnotX[\ii] -\min_i  \EnotX \to \infty$). Thus,  with a sufficiently large gap, any transformation $\ptpp$ can be carried out (as long as $\ps$ and $\pf$ have full support).

Finally, as mentioned in the Introduction, one can generalize the results in the section to account for  constraints on both the energy functions and the non-conservative forces which can be applied to a system. For details, see \cref{app:nonconservative}.

\subsection{Example}
\label{sec:miexample}

We now use \cref{thm:a} to analyze the examples mentioned in the introduction, which involve bringing a system containing  two subsystems $X\times Y$ from an initial correlated distribution to a  final uncorrelated product distribution,  or vice versa, while using decoupled energy functions as in \cref{eq:additiveE}. 
Suppose that the available energy functions only allow the manipulation
of the energy of the up-state of spin $x$ from $-10$ to $10$, while other states are at energy 0. Formally, this corresponds to a line segment of energy functions,
$$\EE=\{\lambda \Enot : \lambda \in[-1,1]\}\text{\quad for\quad} \EnotX[x,y]=10\cdot \delta_{x,1}, $$
which is a special case of \cref{eq:additiveE}.  
Plugging into \cref{eq:eucCap} and rearranging gives %
\begin{align}
\maxCap= 10\cdot \min_{x,y} \max_{x',y'} \vert \delta_{x,1} - \delta_{x',1}\vert.
\label{eq:dddd1}
\end{align}
For any $x,y$,  we have $\max_{x',y'} \vert \delta_{x,1} - \delta_{x',1}\vert=\vert \delta_{x,1} - \delta_{-x,1}\vert =1$, so \cref{eq:dddd1} implies that $\maxCap=10$. 
Then, by \cref{eq:cond1} and \cref{thm:a}, %
any transformation where all initial and final probability values
are greater than $e^{-10\beta}$ is possible. 

Imagine that one wanted to transform a correlated initial distribution $p$ to an uncorrelated final distribution $p'$.
For concreteness, assume that the inverse temperature is $\beta=1$, that the initial distribution is
\begin{align}
\pxs[x,y]=\begin{cases}
0.4 &\text{if $(x,y)\in\{(-1,-1),(1,1)\}$}\\
0.1 &\text{if $(x,y)\in\{(1,-1),(-1,1)\}$}
\end{cases},
\label{eq:corrdist}
\end{align}
and that the final distribution is the uniform $\pxf[x,y]=1/4$. Note that the initial and final mutual information terms obey
$I_{\ps}(X;Y)\approx0.18$ nats and $I_{\pf}(X;Y)=0$. It is also straightforward to verify that the entries of $p$ and $\pf$ are greater than $e^{-10}$, so \cref{eq:cond1} is satisfied. Then, the transformation $\ptpp$ can be implemented using the following procedure: %

\vspace{5pt}

\noindent 1) Choose state $(-1,-1)$ as the ``buffer state''. Then, transfer 0.15 probability from state $(1,1)$ to state $(-1,-1)$ over time $t\in[0,\ft]$;

\vspace{5pt}

\noindent 2) Move 0.15 probability from state $(-1,-1)$ to state $(-1,1)$ via two transfers: first from state $(-1,-1)$ to state $(1,-1)$ over time $t\in[\ft,2\ft]$, then from state $(1,-1)$ to state $(-1,1)$ over time $t\in[2\ft,3\ft]$;

\vspace{5pt}

\noindent 3) Transfer 0.15 probability from state $(-1,-1)$ to state $(1,-1)$ over time $t\in[3\ft,4\ft]$.

\vspace{5pt}

After running this procedure on the initial distribution $\ps$, the system will be left in the uniform distribution $\pf$. Moreover, by taking $\ft$ large enough and using the construction that appears in \cref{app:transfer}, each transfer can be done for an arbitrarily small amount of EP. 

Conversely, imagine that one wanted to increase mutual information by transforming an uncorrelated uniform initial distribution $p$ to a correlated final distribution $p'$. %
Suppose again that $\beta=1$, that $p$ is the uniform distribution, and that $p'$ is given by the right hand side of \cref{eq:corrdist}. Then, this transformation can be accomplished by running the same sequence of transfers as described above, but now in reverse: first transfer 0.15 probability $(1,-1)\to(-1,-1)$, then move 0.15 probability using two transfers $(-1,1)\to(1,-1)$ and $(1,-1)\to(-1,-1)$, and finally transfer 0.15 probability $(-1,-1)\to(1,1)$.

It is important to note that each of the above transfers involves a pair of states which differ in the state of spin $X$, and therefore have a controllable energy gap equal to 10. On the other hand, it is impossible to use a transfer %
to move probability between a pair of states which only differ in the state of spin $Y$, since the corresponding  controllable energy gap will be 0 and \cref{eq:cg2} cannot be satisfied.

\subsection{Physical assumptions behind \cref{thm:a}}
\label{sec:physicalinterpretation}

We briefly discuss three important physical assumptions that underlie the results described above. %

First, we assume that one has access to arbitrarily long driving protocols, i.e., the \emph{quasistatic limit}. This assumption is required for any protocol that transforms some initial distribution $p$ to a different final distribution $p'$ while achieving arbitrarily small EP (as known from research on finite-time thermodynamics~\cite{shiraishi_speed_2018,aurell2011optimal}).  Interestingly, if one does not require that EP is arbitrarily small, then 
arbitrarily long timescales are {not} necessary for carrying out the transformation $\ptpp$ with complete accuracy while using a limited set of energy functions. Instead, as can be seen  from the proof of \cref{lem:transfer} in \cref{app:transfer},  as long as \cref{eq:dyn0,eq:cg2} hold, then 
there is a protocol $\prot=\protSet$ with a finite $\ft$ 
that carries out the transfer $\ptpp$ exactly, while obeying $E(t)\in\EE$ at all $t$. 

Second, we assume that one can impose a \emph{separation of timescales} in which the transition rates between all states are set to zero,  except for a given pair of states involved in a transfer.  
To understand the physical meaning of this assumption, note that setting a transition rate $W_{ji}$ to zero is equivalent to setting the corresponding 
activity parameter $\psi_{ji}$ in \cref{eq:paramRates} to zero. Furthermore, in statistical physics, discrete-state master equations are typically
(though often implicitly) derived by coarse-graining a stochastic
dynamical system over a microscopic continuous phase space, as might be described
by Fokker-Planck dynamics~\citep{talknerRateDescriptionFokkerPlanck2004,van1992stochastic,hanggiReactionrateTheoryFifty1990}. Under this scheme, each coarse-grained state
$\ii$ represents a free energy well in the underlying phase space,  
the corresponding energy level $\Ex[\ii]$ reflects the depth of the free energy well, and the activity parameters $\psi_{ji}$ reflect the height of the free energy barriers separating wells $i$ and $j$. For instance, transition rates are often expressed  using the Arrhenius form $\WjiF= e^{-\beta[B_{ji}-\Ex[\ii]]}$~\citep{astumian2007adiabatic,rahav2008directed,van1992stochastic},  where $B_{\jj \ii }$ is the absolute  height of the free energy barrier separating coarse-grained states $i$ and $j$. This expression can be put in the form of \cref{eq:paramRates} by defining $\psi_{ji}=e^{\beta[(\Ex[\ii]+\Ex[\jj])/2-B_{ji}]}$, so that the activity parameter $\psi_{ji}$ is determined by the {relative} height of the barrier separating coarse-grained states $i$ and $j$.   
From this point of view, our second assumption means that one can impose infinite-sized barriers between all pairs of coarse-grained states, except for a given pair of  states involved in a transfer.

Because we allow certain transition rates to be set to zero, the rate matrices $W(t)$ involved in our driving protocols (as constructed in \cref{app:transfer})  will generally be reducible. For this reason, these rate matrices will have multiple equilibrium distributions (i.e., multiple distributions which are stationary and incur zero EP rate), beyond the unique Boltzmann equilibrium distribution  prescribed by the energy function $E(t)$. In fact, by using such reducible rate matrices, our construction carries out a transformation $\ptpp$ while keeping the system (arbitrary close to) equilibrium throughout, even though those equilibria may be outside the set of the Boltzmann equilibrium distributions that correspond to the energy functions in $\EE$.

Our third assumption  is that any pair of states can be connected by a non-zero transition rate even while the other rates are set to zero (which is a kind of  ``converse'' of our second assumption). 
If, as described above, the discrete-state master equation is derived by coarse-graining a microscopic continuous phase space, then (as above) the validity of this assumption depends on one's ability to manipulate free energy barriers at the microscopic level. In addition, it also depends on the geometric properties of the embedding of the coarse-grained states within the microscopic space (e.g., if the underlying microscopic continuous space is one-dimensional, then not all patterns of connectivity between states can be achieved by manipulating free energy barriers). 
It is possible to generalize our treatment to account for constraints on which states can/cannot be connected via non-zero transition rates (e.g., by defining $\egS=0$ for any pair of states $i,j$ that cannot be connected by a non-zero transition rate), though for simplicity we do not consider this generalization in the current paper.

\section{EP under both constraints on energy functions and rate matrices}
\label{sec:second}

In real-world scenarios,  %
there are often additional restrictions on the transition rates, not only on the energy function, which can preclude the use of the  protocols
constructed in the previous section. For instance, such restrictions might arise   because there are 
constraints on the how the heights of the
free energy barriers between coarse-grained states  can be manipulated, meaning that some transition rates cannot be set to 0. 
 Accordingly, in this section we derive our second main result, which is a 
bound on the EP that arises under more general constraints on the rate matrices. 

As before, we consider a system coupled to a single heat bath at inverse temperature $\beta$. %
Suppose that one drives the system from some initial distribution $\ps$ to some final distribution $\pf$, while only using energy functions in some restricted set $\EE$ and rate matrices in some restricted set $\RR$.  
We assume that there exists \emph{some} driving protocol $\prot = \protSet$ 
that implements the desired transformation $\ptpp$, while obeying $E(t)\in\EE$ and $\R(t)\in\RR$ at all $t$. We then analyze the minimal EP that must be incurred by this protocol in terms of the properties of $\EE$ and $\RR$. 
Note that in this section, we do not assume that either $\EE$ or $\RR$ is path-connected (in fact, we will sometimes assume that both $\EE$ and $\RR$ are finite sets). 

Before presenting our result, we introduce the notion of a ``KL projection'' of a distribution $p$ onto the set of available equilibrium distributions, which we indicate as $\klproj$~\cite{csiszar2003information}. The KL projection of $p$ refers to the Boltzmann equilibrium distribution that is closest to $p$ in terms of KL divergence, as defined in \cref{eq:kldef}, among the set of all Boltzmann distribution allowed by the available energy functions: %
\begin{equation}
\klproj:=\argmin_{E\in\EE}\DKL(p\Vert\piE).\label{eq:klproj}
\end{equation}
(As above, we use the notation  $\piE = e^{-\beta E}/Z$ to indicate the Boltzmann distribution
corresponding to energy function $E$.) 
Note that if  $p$ is a Boltzmann distribution for some $E\in\EE$, then $\klproj = p$.

Next, suppose that there is some number $\mult \in [0,1]$
such that %
\begin{align}
\EPr(p,W) \ge-\mult\sum_{\ii,\jj} \px \WjiF \ln\frac{\px[\jj]}{[\klproj]_\jj}\quad \forall p,\R\in\RR,\label{eq:optA}
\end{align}
where $\EPr(p,W)$ is the rate of EP incurred by rate matrix $W$ on distribution $p$, as defined in \cref{eq:eprate0}.  
This is an implicit inequality that bounds $\mult$ in terms of the set  of 
allowed rate matrices $\RR$ as well as  the KL projection $\klproj$, which in turn is a function of the allowed energy functions $\EE$. 
This inequality is always satisfied  for $\mult = 0$, since  it is then equivalent to the statement that the EP rate is non-negative. However, we will also consider sets $\EE$ and $\RR$ where \cref{eq:optA} holds for some $\mult>0$.  
Importantly, as long as the set of allowed rate matrices $\RR$ is
finite, it is possible to find the largest value of $\mult$ that
satisfies \cref{eq:optA} via numerical convex optimization techniques.
The procedure for doing this is described in detail in \cref{app:convexopt},
and code is provided at \url{https://github.com/artemyk/epbound}.

The right hand side of \cref{eq:optA}
is $\mult$ times the rate at which the distribution $p$ approaches
the closest equilibrium distribution $\klproj$ under the dynamics generated by $\R$ (see \cref{eq:z0} in \cref{app:eprateEtaBound}). Thus, the multiplier $\mult$ bounds how much faster any
distribution $p$ can approach the closest equilibrium $\klproj$,
relative to the rate at which that distribution incurs EP. 
Note also that 
$\mult$ is independent of the overall timescale of the rate
matrices $\R\in\RR$: if \cref{eq:optA} holds for
some $\mult$, some distribution $p$, and some rate matrix $\R$, then
it will also hold when that rate matrix is rescaled as $\R \to \lambda\R$
(since this is equivalent to multiplying both sides 
by $\lambda$).

Now consider  the total EP incurred by the protocol $\prot$ on initial distribution $p$, as defined in \cref{eq:EPdefInt}.  Using  \cref{eq:optA} and some simple rearrangement, the EP rate at time $t$ can be bound in terms of the time derivative of the KL divergence between $p(t)$ (the distribution at time $t$) and its KL projection $\klprojT$, 
\begin{align}
\EPr(p(t),\R(t))\ge-\mult\ddt \DKL(\pt\Vert\klprojT).
\label{eq:ineq34}
\end{align}
(See \cref{app:eprateEtaBound} for details.) 
This leads to our second main result, which follows from integrating both sides of \cref{eq:ineq34} and using the fundamental theorem of calculus.

\begin{thm}
\label{thm:2}
Let $\prot=\protSet$ be a protocol such that $\Et\in\EE$ and $\Rt \in\RR$ at all $t$. Then, %
\[
\EP(p,\prot)\ge\mult\big[\DKL(\ps\Vert\klprojPS)-\DKL(\pf\Vert\klprojPF)\big]
\]
for  any $\mult\in[0,1]$ where \cref{eq:optA} holds, and $\klproj[\cdot]$ as in \cref{eq:klproj}. 
\end{thm}
\noindent

The EP bound in \cref{thm:2} is the product of two terms: the drop
of the KL divergence $D(p\Vert\klproj)$ 
and a ``multiplier'' $\mult$. %
The drop of KL divergence reflects the contribution to EP arising
from constraints on the energy functions, as determined by the set $\EE$. In particular, if all energy
functions are available, then $\klproj=p$ and this KL divergence
term vanishes, leading to a trivial bound $\EP\ge0$.  
The multiplier $\mult$ reflects the contribution to the EP bound that arises from constraints on the rate matrices, as determined both by $\EE$ and $\RR$. 
Note that 
\cref{thm:2} only gives non-trivial bounds on EP when the final distribution
is closer to the set of equilibrium distributions than the initial
one, so that $D(\ps\Vert\klprojPS) > D(\pf\Vert\klprojPF)$.

Interestingly, our result holds not only when $\klproj$ is defined as the KL projection to the set of equilibrium distributions, as in \cref{eq:klproj}, but more generally when $\klproj$ is defined as the KL projection to \emph{any} arbitrary set of distributions $\distSet$. In other words, 
if one defines $\klproj=\argmin_{q\in\distSet}D(p\Vert q)$ for any   $\distSet$ and then finds a corresponding $\mult\in[0,1]$ such that \cref{eq:optA} holds, then  
\cref{thm:2} will still apply. 
Each set of distributions $\distSet$ will have its own maximal value of $\mult$ which satisfies \cref{eq:optA} and  will therefore induce its bound on EP. It may be possible to derive tighter bounds on EP in  \cref{thm:2} by varying the choice of the distributions $\distSet$ in the definition of $\klproj$, though we leave exploration of this choice for future work.

In fact, by exploiting the freedom in how the KL projection $\klproj$ can be defined, it is possible to generalize \cref{thm:2} to derive bounds on EP in the presence of non-conservative forces and multiple thermodynamic reservoirs (rather than a single heat bath as considered above). Details of this generalization are discussed in \cref{app:thm2multiplebaths}.

\begin{figure}
\begin{centering}
\includegraphics[width=0.4\columnwidth]{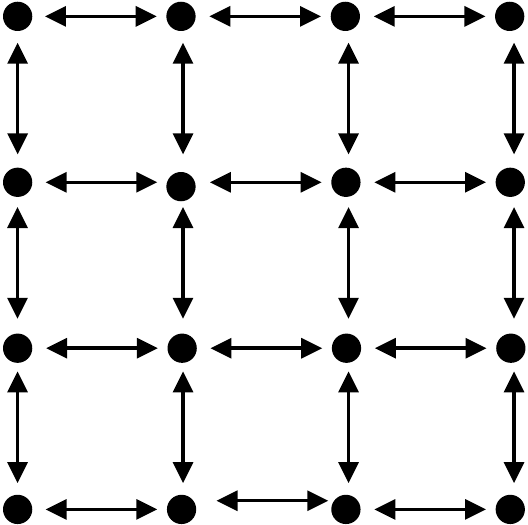}
\par\end{centering}
\caption{A discrete-state system with 16 states arranged in a
4-by-4 lattice, with only nearest neighbor transitions allowed. $\protect\RR$
contains 16 rate matrices, each one corresponding to an equilibrium
distribution in which the energy of a particular state is increased.\label{fig:opt}}
\end{figure}

\newcommand{\ElocI}{E^{(\ii)}}
\newcommand{\PilocI}{\pi^{(\ii)}}
\newcommand{\RlocI}{\R^{(\ii)}}

\subsection{Example}

We demonstrate \cref{thm:2} with a discretized model of a Brownian particle on a 2-dimensional
lattice, where we imagine that one can only increase the energy of a single lattice site at any one time. 
Assume that the system is coupled to a heat bath at inverse temperature $\beta$, and that each system state $\ii$ is identified with
a location on a 2-dimensional $N\times N$
lattice, as in \cref{fig:opt}. Suppose that there are $N^{2}$
rate matrices in $\RR$ and $N^{2}$ energy functions in $\EE$, one
for each location. The energy function corresponding to location $\ii$ assigns energy 1 to state $\ii$ and energy
$0$ to all other states, 
 $\ElocI_\jj=\delta_{\ii,\jj}$. 
The rate matrix corresponding to state
$\ii$ allows only nearest-neighbor
transitions: the off-diagonal entries obey $\RlocI_{kj}=e^{\beta \ElocI_j}$
when states $j$ and $k$ are nearest neighbors on the 2-dimensional
lattice, and otherwise $\RlocI_{kj}=0$. 

We now use  \cref{thm:2} to derive a bound on EP for this system. For concreteness, we take $\beta=1$ and $N=4$. Using the convex optimization technique described in \cref{app:convexopt}, we find that the tightest bound in \cref{eq:optA} is given by $\mult\approx0.6$. Thus, for any transformation $\ptpp$, %
\begin{equation}
\EP\ge0.6\left[D(\ps\Vert\klprojPS)-D(\pf\Vert\klprojPF)\right].\label{eq:ex9b}
\end{equation}
Given the structure of the available energy functions, %
the KL divergence to any equilibrium distribution $\pi^{\ElocI}=e^{-\ElocI}/Z$ can be written as 
\[
D(p\Vert\pi^{\ElocI})= p_\ii+\ln Z-S(p),
\]
where $Z=\sum_{\jj}e^{-\delta_{\ii,\jj}}=(N^{2}-1)+e^{-1}$
 is a normalization constant. %
Thus, the KL divergence to the closest equilibrium distribution can be written as $D(p\Vert\klproj)=\min_{\ii}p_\ii+\ln Z-S(p)$. Plugged
into \cref{eq:ex9b}, this gives the bound 
\[
\EP\ge0.6\big[S(\pf)-S(\ps)+(\min_{\ii}\pxs[\ii]-\min_{\ii}\pxf[\ii])\big].
\]
In words, for this set of constrained driving protocols, at least
$0.6$ of the increase in the entropy of the system, plus the drop in the minimum probability value in going from $\ps$
to $\pf$, must be dissipated as EP.

\section{Discussion and future work}

In this paper, we considered a finite-state classical system that undergoes a driving protocol, represented as a time-dependent master equation. We derived two results concerning the feasibility of transforming such a system from some initial distribution $\ps$ to a final distribution $\pf$ given constraints on the driving protocol, and the associated EP.

Our first result, presented in \cref{sec:first}, concerned the case where the driving protocol can only access a restricted set of energy functions $\EE$, but can otherwise make use of arbitrary rate matrices (as long as they obey local detailed balance).  We showed that any given transformation $\ptpp$ can be implemented for a vanishing amount of EP,
so long as the probabilities in $\ps$ and $\pf$ are bounded away from zero by an amount which depends on $\EE$. We also demonstrated that this condition is quite weak, since even a limited amount of control over the energy function (such as the ability to manipulate the energy of a single state) suffices to implement an arbitrary transformation with vanishing EP.  This result is derived under the assumption that one can control all other aspects of the rate matrices beyond the energy function, including the ability to set arbitrary transition rates to zero/non-zero values.

In our second result, presented in \cref{sec:second}, we derived a lower bound on the EP involved in carrying out some transformation $\ptpp$ in the presence of constraints on both the energy functions and the rate matrices. We show that in some cases, this bound can be determined numerically using standard convex optimization techniques. 

We briefly mention some possible directions for future work. First, in deriving our first result, we do not prove that  \cref{eq:cond1} is %
a \emph{necessary} condition for carrying out the transformation $\ptpp$ with vanishing EP while using energy functions in $\EE$, only that it is a \emph{sufficient} condition. An interesting research direction for future work would investigate sufficient {and} necessary conditions for carrying out  such a transformation.

Second, in deriving our first result, we constructed a protocol that carries out $\ptpp$ without restricting how the protocol behaves on other initial distributions besides $\ps$. Future work may consider the related, but more difficult, problem of carrying out a \emph{logical map} (a conditional probability distribution) $T_{\ii\jj}$ from initial states $\jj$ to final states $\ii$ while using a limited set of energy functions. In other words, this research direction would analyze  driving protocols which  {(1)} implement some desired logical map $T_{\ii\jj}$, {(2)} achieve vanishing EP on some particular initial  distribution $\ps$~\cite{kolchinsky2016dependence}, and {(3)} only use energy functions from some limited set $\EE$.  
It is possible that this problem can be tackled by combining the ``transfer'' protocol used in \cref{lem:transfer} with the constructions developed in \cite{owen_number_2018,wolpert2019space}, which show how to implement a given logical map with continuous-time master equations (while possibly using some number of auxiliary ``hidden'' states). 

Third, in this paper we considered finite-state master equation dynamics. Future work may investigate whether our results can be extended to  continuous-state dynamics, such as Fokker-Planck dynamics over probability densities. Such an extension would be far from trivial, as our first result relies on the ability to set arbitrary transition rates to zero/non-zero values (which is inappropriate for Fokker-Planck dynamics, even when discretized). 
On the other hand, our second result can be formally generalized to continuous-state dynamics, but it is not clear when the corresponding infinite-dimensional optimization problem (as defined in \cref{app:convexopt}) can be solved in practice. We note that in previous work~\cite{kolchinsky2020entropy}, we investigated bounds on EP for Fokker-Planck dynamics under highly structured constraints on the energy functions, such as symmetry, modularity, and coarse-graining constraints.

Finally, future work may consider whether the methods developed here can be extended to Markovian quantum systems. This research direction would analyze an open quantum system evolving according to time-inhomogeneous Lindbladian dynamics~\cite{spohn_entropy_1978}, and investigate how constraints on the available Hamiltonians and Lindbladian operators translate into bounds on the quantum EP involved in bringing the system from some initial mixed state $\rho$ to some final mixed state $\rho'$.

\section{Acknowledgments}

We thank Massimiliano Esposito, Alec
Boyd, and the anonymous reviewers for helpful discussions and suggestions. %
This research was supported by grant
number FQXi-RFP-IPW-1912 from the Foundational Questions Institute
and Fetzer Franklin Fund, a donor advised fund of Silicon Valley Community
Foundation. The authors thank the Santa Fe Institute for helping to
support this research.

\noindent 
\bibliographystyle{IEEEtran}
\bibliography{main}

\begin{thebibliography}{10}
\providecommand{\url}[1]{#1}
\csname url@samestyle\endcsname
\providecommand{\newblock}{\relax}
\providecommand{\bibinfo}[2]{#2}
\providecommand{\BIBentrySTDinterwordspacing}{\spaceskip=0pt\relax}
\providecommand{\BIBentryALTinterwordstretchfactor}{4}
\providecommand{\BIBentryALTinterwordspacing}{\spaceskip=\fontdimen2\font plus
\BIBentryALTinterwordstretchfactor\fontdimen3\font minus
  \fontdimen4\font\relax}
\providecommand{\BIBforeignlanguage}[2]{{%
\expandafter\ifx\csname l@#1\endcsname\relax
\typeout{** WARNING: IEEEtran.bst: No hyphenation pattern has been}%
\typeout{** loaded for the language `#1'. Using the pattern for}%
\typeout{** the default language instead.}%
\else
\language=\csname l@#1\endcsname
\fi
#2}}
\providecommand{\BIBdecl}{\relax}
\BIBdecl

\bibitem{seifert2012stochastic}
U.~Seifert, ``Stochastic thermodynamics, fluctuation theorems and molecular
  machines,'' \emph{Reports on Progress in Physics}, vol.~75, no.~12, p.
  126001, 2012.

\bibitem{esposito2011second}
M.~Esposito and C.~Van~den Broeck, ``Second law and {L}andauer principle far
  from equilibrium,'' \emph{EPL (Europhysics Letters)}, vol.~95, no.~4, p.
  40004, 2011.

\bibitem{takara_generalization_2010}
K.~Takara, H.-H. Hasegawa, and D.~Driebe, ``Generalization of the second law
  for a transition between nonequilibrium states,'' \emph{Physics Letters A},
  vol. 375, no.~2, pp. 88--92, 2010.

\bibitem{parrondo2015thermodynamics}
J.~M. Parrondo, J.~M. Horowitz, and T.~Sagawa, ``Thermodynamics of
  information,'' \emph{Nature Physics}, vol.~11, no.~2, pp. 131--139, 2015.

\bibitem{van2013stochastic}
C.~Van~den Broeck, ``Stochastic thermodynamics: A brief introduction,''
  \emph{Phys. Complex Colloids}, vol. 184, pp. 155--193, 2013.

\bibitem{maes2018non}
C.~Maes, \emph{Non-dissipative effects in nonequilibrium systems}.\hskip 1em
  plus 0.5em minus 0.4em\relax Springer, 2018.

\bibitem{Note1}
It is known that for a free relaxation from an initial distribution $\ps $ to a
  final equilibrium distribution $\pf =\piE $, the EP obeys $\EP =\DKL (\ps
  \Vert \piE )$~\cite {schloglStatisticalFoundationThermodynamic1967}. In
  addition, we can rewrite $\DKL (\ps \Vert \piE ) = I_{\ps }(X;Y) + \DKL (p_X
  p_Y \Vert \piE _X \piE _Y) \ge I_{\ps }(X;Y)$. Combining implies $\EP \ge
  I_{\ps }(X;Y)$.

\bibitem{wolpert2020uncertainty}
D.~H. Wolpert, ``Uncertainty relations and fluctuation theorems for bayes
  nets,'' \emph{Physical Review Letters}, vol. 125, no.~20, p. 200602, 2020.

\bibitem{wolpert2020minimal}
D.~Wolpert, ``Minimal entropy production rate of interacting systems,''
  \emph{New Journal of Physics}, 2020.

\bibitem{wolpert2019stochastic}
D.~H. Wolpert, ``The stochastic thermodynamics of computation,'' \emph{Journal
  of Physics A: Mathematical and Theoretical}, vol.~52, no.~19, p. 193001,
  2019.

\bibitem{wolpert2020thermodynamic}
D.~H. Wolpert and A.~Kolchinsky, ``Thermodynamics of computing with circuits,''
  \emph{New Journal of Physics}, 2020.

\bibitem{boyd2018thermodynamics}
A.~B. Boyd, D.~Mandal, and J.~P. Crutchfield, ``Thermodynamics of modularity:
  Structural costs beyond the landauer bound,'' \emph{Physical Review X},
  vol.~8, no.~3, p. 031036, 2018.

\bibitem{kolchinsky2020entropy}
A.~Kolchinsky and D.~H. Wolpert, ``Work, entropy production, and thermodynamics
  of information under protocol constraints,'' \emph{arXiv preprint
  arXiv:2008.10764}, 2020.

\bibitem{wilming_second_2016}
H.~Wilming, R.~Gallego, and J.~Eisert, ``Second law of thermodynamics under
  control restrictions,'' \emph{Physical Review E}, vol.~93, no.~4, 2016.

\bibitem{lostaglio2018elementary}
M.~Lostaglio, {\'A}.~M. Alhambra, and C.~Perry, ``Elementary thermal
  operations,'' \emph{Quantum}, vol.~2, p.~52, 2018.

\bibitem{baumerImperfectThermalizationsAllow2019}
E.~B{\"a}umer, M.~{Perarnau-Llobet}, P.~Kammerlander, H.~Wilming, and
  R.~Renner, ``Imperfect {{Thermalizations Allow}} for {{Optimal Thermodynamic
  Processes}},'' \emph{Quantum}, vol.~3, p. 153, 2019.

\bibitem{perrySufficientSetExperimentally2018}
C.~Perry, P.~{\'C}wikli{\'n}ski, J.~Anders, M.~Horodecki, and J.~Oppenheim, ``A
  {{Sufficient Set}} of {{Experimentally Implementable Thermal Operations}} for
  {{Small Systems}},'' \emph{Physical Review X}, vol.~8, no.~4, p. 041049,
  2018.

\bibitem{owen_number_2018}
J.~A. Owen, A.~Kolchinsky, and D.~H. Wolpert, ``Number of hidden states needed
  to physically implement a given conditional distribution,'' \emph{New Journal
  of Physics}, 2018.

\bibitem{wolpert2019space}
D.~H. Wolpert, A.~Kolchinsky, and J.~A. Owen, ``A space--time tradeoff for
  implementing a function with master equation dynamics,'' \emph{Nature
  communications}, vol.~10, no.~1, p. 1727, 2019.

\bibitem{korzekwa2021quantum}
K.~Korzekwa and M.~Lostaglio, ``Quantum advantage in simulating stochastic
  processes,'' \emph{Physical Review X}, vol.~11, no.~2, p. 021019, 2021.

\bibitem{esposito2010finite}
M.~Esposito, R.~Kawai, K.~Lindenberg, and C.~Van~den Broeck, ``Finite-time
  thermodynamics for a single-level quantum dot,'' \emph{EPL (Europhysics
  Letters)}, vol.~89, no.~2, p. 20003, 2010.

\bibitem{sivak2012thermodynamic}
D.~A. Sivak and G.~E. Crooks, ``Thermodynamic metrics and optimal paths,''
  \emph{Physical Review Letters}, vol. 108, no.~19, p. 190602, 2012.

\bibitem{shiraishi_speed_2018}
N.~Shiraishi, K.~Funo, and K.~Saito, ``Speed limit for classical stochastic
  processes,'' \emph{Physical Review Letters}, vol. 121, no.~7, 2018.

\bibitem{gomez2008optimal}
A.~Gomez-Marin, T.~Schmiedl, and U.~Seifert, ``Optimal protocols for minimal
  work processes in underdamped stochastic thermodynamics,'' \emph{The Journal
  of chemical physics}, vol. 129, no.~2, p. 024114, 2008.

\bibitem{then2008computing}
H.~Then and A.~Engel, ``Computing the optimal protocol for finite-time
  processes in stochastic thermodynamics,'' \emph{Physical Review E}, vol.~77,
  no.~4, p. 041105, 2008.

\bibitem{zulkowski2014optimal}
P.~R. Zulkowski and M.~R. DeWeese, ``Optimal finite-time erasure of a classical
  bit,'' \emph{Physical Review E}, vol.~89, no.~5, p. 052140, 2014.

\bibitem{schmiedl2007optimal}
T.~Schmiedl and U.~Seifert, ``Optimal finite-time processes in stochastic
  thermodynamics,'' \emph{Physical Review Letters}, vol.~98, no.~10, p. 108301,
  2007.

\bibitem{remlein2021optimality}
B.~Remlein and U.~Seifert, ``Optimality of nonconservative driving for
  finite-time processes with discrete states,'' \emph{Physical Review E}, vol.
  103, no.~5, p. L050105, 2021.

\bibitem{aurell2011optimal}
E.~Aurell, C.~Mej{\'\i}a-Monasterio, and P.~Muratore-Ginanneschi, ``Optimal
  protocols and optimal transport in stochastic thermodynamics,''
  \emph{Physical Review Letters}, vol. 106, no.~25, p. 250601, 2011.

\bibitem{aurell2012refined}
E.~Aurell, K.~Gaw\c{e`}dzki, C.~Mej{\'\i}a-Monasterio, R.~Mohayaee, and
  P.~Muratore-Ginanneschi, ``Refined second law of thermodynamics for fast
  random processes,'' \emph{Journal of statistical physics}, vol. 147, no.~3,
  pp. 487--505, 2012.

\bibitem{machta2015dissipation}
B.~B. Machta, ``Dissipation bound for thermodynamic control,'' \emph{Physical
  Review Letters}, vol. 115, no.~26, p. 260603, 2015.

\bibitem{verley_work_2014}
G.~Verley, C.~V.~d. Broeck, and M.~Esposito, ``Work statistics in
  stochastically driven systems,'' \emph{New Journal of Physics}, vol.~16,
  no.~9, p. 095001, 2014.

\bibitem{gour2015resource}
G.~Gour, M.~P. M{\"u}ller, V.~Narasimhachar, R.~W. Spekkens, and N.~Y. Halpern,
  ``The resource theory of informational nonequilibrium in thermodynamics,''
  \emph{Physics Reports}, vol. 583, pp. 1--58, 2015.

\bibitem{schnakenberg1976network}
J.~Schnakenberg, ``Network theory of microscopic and macroscopic behavior of
  master equation systems,'' \emph{Reviews of Modern physics}, vol.~48, no.~4,
  p. 571, 1976.

\bibitem{esposito2010three}
M.~Esposito and C.~Van~den Broeck, ``Three faces of the second law. {I}.
  {M}aster equation formulation,'' \emph{Physical Review E}, vol.~82, no.~1, p.
  011143, 2010.

\bibitem{Note2}
There are at most $n-1$ states that have either $\pxs <\pxf $ or $\pxs >\pxf $
  and are not state 1. Furthermore, each path can have at most $\ell =n-1$
  steps. Thus, the number of transfers involved in the two-stage construction
  is no greater than $(n-1)^2$.

\bibitem{pollack1960letter}
M.~Pollack, ``Letter to the editor--the maximum capacity through a network,''
  \emph{Operations Research}, vol.~8, no.~5, pp. 733--736, 1960.

\bibitem{hu1961letter}
T.~Hu, ``Letter to the editor--the maximum capacity route problem,''
  \emph{Operations Research}, vol.~9, no.~6, pp. 898--900, 1961.

\bibitem{talknerRateDescriptionFokkerPlanck2004}
P.~Talkner and J.~{\L}uczka, ``Rate description of {{Fokker}}-{{Planck}}
  processes with time-dependent parameters,'' \emph{Physical Review E},
  vol.~69, no.~4, 2004.

\bibitem{van1992stochastic}
N.~G. Van~Kampen, \emph{Stochastic processes in physics and chemistry}.\hskip
  1em plus 0.5em minus 0.4em\relax Elsevier, 1992, vol.~1.

\bibitem{hanggiReactionrateTheoryFifty1990}
P.~H{\"a}nggi, P.~Talkner, and M.~Borkovec, ``Reaction-rate theory: Fifty years
  after {{Kramers}},'' \emph{Reviews of Modern Physics}, vol.~62, no.~2, pp.
  251--341, 1990.

\bibitem{astumian2007adiabatic}
R.~D. Astumian, ``Adiabatic operation of a molecular machine,''
  \emph{Proceedings of the National Academy of Sciences}, vol. 104, no.~50, pp.
  19\,715--19\,718, 2007.

\bibitem{rahav2008directed}
S.~Rahav, J.~Horowitz, and C.~Jarzynski, ``Directed flow in nonadiabatic
  stochastic pumps,'' \emph{Physical Review Letters}, vol. 101, no.~14, p.
  140602, 2008.

\bibitem{csiszar2003information}
I.~Csisz{\'a}r and F.~Matus, ``Information projections revisited,'' \emph{IEEE
  Transactions on Information Theory}, vol.~49, no.~6, pp. 1474--1490, 2003.

\bibitem{kolchinsky2016dependence}
A.~Kolchinsky and D.~H. Wolpert, ``Dependence of dissipation on the initial
  distribution over states,'' \emph{Journal of Statistical Mechanics: Theory
  and Experiment}, p. 083202, 2017.

\bibitem{spohn_entropy_1978}
H.~Spohn, ``Entropy production for quantum dynamical semigroups,''
  \emph{Journal of Mathematical Physics}, vol.~19, no.~5, pp. 1227--1230, 1978.

\bibitem{schloglStatisticalFoundationThermodynamic1967}
F.~Schl{\"o}gl, ``On the statistical foundation of the thermodynamic evolution
  criterion of {{Glansdorff}} and {{Prigogine}},'' \emph{Annals of Physics},
  vol.~45, no.~1, pp. 155--163, 1967.

\bibitem{maes2013minimum}
C.~Maes and K.~Netocny, ``Minimum entropy production principle,''
  \emph{Scholarpedia}, vol.~8, no.~7, p. 9664, 2013.

\end{thebibliography}

\appendix
\crefalias{section}{appsec}
\crefalias{subsection}{appsubsec}

\section{Proof of \cref{lem:transfer}}

\label{app:transfer}

\global\long\def\protAlpha{\Gamma_{\ft}}%
\global\long\def\Etalpha{E^{\ft}(t)}%
\global\long\def\Ralpha{\R^{\ft}}%
\global\long\def\EPinstTau{\dot{\EP}(p^{\tau}(t),\Ralpha(t))}%

We prove the result by construction. In particular, we first construct a family of protocols $\protAlpha=\{(\Etalpha,\Ralpha(t)):t\in[0,\ft]\}$
parameterized by temporal duration $\ft$ (where  
$\ft\to\infty$ corresponds to the quasistatic limit). We then show that if \cref{eq:dyn0} and \cref{eq:cg2} hold, then for any $\epsilon>0$,
there is some $\ft$ such that the protocol $\protAlpha$ satisfies
the following three conditions:
\begin{enumerate}
\item[I.] The protocol maps the initial distribution $\ps$ at time $t=0$
to the final distribution $p'$ at time $t=\ft$, where
\begin{equation}
\pf_{i}=p_{i}+\Delta(\delta_{b,i}-\delta_{a,i}).\label{eq:n233}
\end{equation}
\item[II.] The EP obeys $\EP(p,\protAlpha)\le\epsilon$.
\item[II.] The energy functions obey $\Etalpha\in\EE$ at all $t$. 
\end{enumerate}

We assume below that $\Delta\ge0$ in \cref{eq:n233}, so that $p_{a}'\le p_{a}$. 
Note that this is done without loss of generality (if $\Delta<0$,
one can swap the roles of states $a$ and $b$).

We begin with a few definitions. First, for any $\ft>0$, define the following function
$f_{\tau}:[0,\ft]\to\mathbb{R}$,
\begin{equation}
f_{\tau}(t):=p_{a}+\frac{t}{\tau-1+e^{-\tau}}(p_{a}'-p_{a}).\label{eq:fdef0}
\end{equation}
Second, let $C=p_{a}+p_{b}=p_{a}'+p_{b}'$ refer to the sum of the
probability in states $a$ and $b$. Third, define a time-dependent
rate matrix $\R^{\tau}$ with the following entries for $i\ne j$:
\begin{align}
\R_{ij}^{\tau}(t) & =\begin{cases}
f_{\tau}(t)/C & \text{if \ensuremath{i=a,j=b}}\\
1-f_{\tau}(t)/C & \text{if \ensuremath{i=b,j=a}}\\
0 & \text{otherwise}
\end{cases}\label{eq:trans2}
\end{align}

Given  \cref{eq:fdef0} and the inequality $\ft-1+e^{-\ft}>0$, as well as the assumption that $p_{a}'\le p_{a}$, it can be verified that  
$f_{\tau}(t)$ is monotonically decreasing in $t$, so
\begin{equation}
p_{a}+\frac{\ft}{\tau-1+e^{-\tau}}(p_{a}'-p_{a})=f_{\tau}(\ft)\le f_{\tau}(t)\le f_{\tau}(0)=p_{a}.\label{eq:bnd2}
\end{equation}
Observe that the lower bound above converges to $p_{a}'$ as $\ft\to\infty$ and that  $p_{a}'>0$, which follows \cref{eq:cg2}. In the following,
we will assume that $\tau$ is large enough so that 
\begin{equation}
f_{\tau}(\ft)\ge p_{a}'/2>0\qquad\forall t\in[0,\ft].\label{eq:bd23}
\end{equation}
Note that given  \cref{eq:bnd2,eq:trans2}, this implies that the
transition rates in $\Ralpha$ are always bounded
between $0$ and $1$.

Let $p^{\ft}(t)$ indicate a time-dependent probability which undergoes
the Markovian evolution $\ddt p^{\ft}(t)=\Ralpha(t)p^{\ft}(t)$ starting
from the initial condition $p^{\ft}(0)=p$. It is clear from \cref{eq:trans2}
that $\ddt p_{i}^{\ft}(t)=0$ for any $i\not\in\{a,b\}$, therefore
the sum of the probability of states $a,b$ is conserved over time,
\begin{equation}
p_{a}^{\ft}(t)+p_{b}^{\ft}(t)=C\qquad\forall t\in[0,\ft].\label{eq:gd32}
\end{equation}
Next, write the temporal derivative of $p_{a}^{\ft}(t)$ as
\begin{align}
\ddt p_{a}^{\ft}(t) & =\R_{ab}^{\ft}(t)p_{b}^{\ft}(t)-\R_{ba}^{\ft}(t)p_{a}^{\ft}(t)\nonumber \\
 & =(f_{\ft}(t)/C)(C-p_{a}^{\ft}(t))-(1-f_{\ft}(t)/C)p_{a}^{\ft}(t)\nonumber \\
 & =f_{\ft}(t)-p_{a}^{\ft}(t),\label{eq:d2}
\end{align}
where we used \cref{eq:trans2} and \cref{eq:gd32}.
The differential equation in \cref{eq:d2} can be solved using some
calculus (or a symbolic computation package such as Mathematica) to give
\begin{align}
p_{a}^{\ft}(t) & =\left(1-\frac{t-1+e^{-t}}{\ft-1+e^{-\ft}}\right)p_{a}+\frac{t-1+e^{-t}}{\ft-1+e^{-\tau}}p'_{a}.\label{eq:sol3}
\end{align}

We now demonstrate that the protocol constructed above satisfies conditions
I-III.

We begin with condition I. From \cref{eq:sol3}, it can be verified by inspection 
that $p_{a}^{\ft}(t)$ evolves from the initial probability $p_{a}^{\ft}(0)=p_{a}$
to the final probability $p_{a}^{\ft}(\ft)=p_{a}'$. In combination
with \cref{eq:gd32} and \cref{eq:n233}, this means that $p_{b}^{\ft}(t)$
evolves from the initial probability $p_{b}^{\ft}(0)=p_{b}$ to the
final probability $p_{b}^{\ft}(\ft)=p_{b}+(p_{a}-p_{a}')=p_{b}'$.
Meanwhile, the probability of all states $i\not\in\{a,b\}$ stays
constant. %

Next, we prove that condition II is satisfied by showing that %
\begin{equation}
\lim_{\ft\to\infty}\EP(p,\protAlpha)=0,\label{eq:lim2}
\end{equation}
which means that EP can be made arbitrarily
small by choosing a sufficiently large $\ft$. First, plug \cref{eq:fdef0} and \cref{eq:sol3} into \cref{eq:d2}
and simplify to give
\begin{align}
\ddt p_{a}^{\ft}(t) & =\R_{ab}^{\ft}(t)p_{b}^{\ft}(t)-\R_{ba}^{\ft}(t)p_{a}^{\ft}(t)\nonumber \\
 & =\frac{1-e^{-t}}{\ft-1+e^{-\ft}}(p_{a}'-p_{a})\le0.\label{eq:hg4}
\end{align}
where the last inequality follows from $p_{a}'\le p_{a}$. 
Next, given   \cref{eq:eprate0},  %
the EP rate incurred by protocol $\protAlpha$ at time $t$ is %
\begin{multline}
\EPinstTau=\\
(\R_{ba}^{\ft}(t)p_{a}^{\ft}(t)-\R_{ab}^{\ft}(t)p_{b}^{\ft}(t))\ln\frac{\R_{ba}^{\ft}(t)p_{a}^{\ft}(t)}{\R_{ab}^{\ft}(t)p_{b}^{\ft}(t)}.\label{eq:gf3}
\end{multline}
Since $\R_{ba}^{\ft}(t)p_{a}^{\ft}(t)\ge\R_{ab}^{\ft}(t)p_{b}^{\ft}(t)$
from \cref{eq:hg4}, \cref{eq:gf3} is the product of two positive terms.
We then use the inequality $\ln x\le(x-1)$ to bound the EP rate as
\begin{multline}
\EPinstTau  \le\frac{(\R_{ba}^{\ft}(t)p_{a}^{\ft}(t)-\R_{ab}^{\ft}(t)p_{b}^{\ft}(t))^{2}}{\R_{ab}^{\ft}(t)p_{b}^{\ft}(t)} \\
 =\frac{(p_{a}'-p_{a})^{2}}{\R_{ab}^{\ft}(t)p_{b}^{\ft}(t)}\Big(\frac{1-e^{-t}}{\ft-1+e^{-\ft}}\Big)^{2},\label{eq:bnd7}
\end{multline}
where in the second line we used \cref{eq:hg4}. Observe that $$\R_{ab}^{\ft}(t)=f_{\ft}(t)/C\ge p_{a}'/(2C)$$
from \cref{eq:trans2,eq:bnd2,eq:bd23}, and that $p_{b}^{\tau}(t)$
is increasing in $t$, since $\ddt p_{b}^{\ft}(t)=-\ddt p_{a}^{\ft}(t)\ge0$ from \cref{eq:hg4}. Thus, %
\[
p_{b}^{\ft}(t)\ge p_{b}^{\ft}(0)=p_{b}>0,
\]
where the last inequality is implied by \cref{eq:cg2}. Plugging these
bounds into \cref{eq:bnd7} gives
\begin{align*}
\EPinstTau & \le\frac{2C(p_{a}'-p_{a})^{2}}{p_{a}'p_{b}}\Big(\frac{1-e^{-t}}{\ft-1+e^{-\ft}}\Big)^{2}\\
&\le\frac{2C(p_{a}'-p_{a})^{2}}{p_{a}'p_{b}}\frac{1}{(\ft-1+e^{-\ft})^2}.
\end{align*}
Using \cref{eq:EPdefInt} and integrating, we can bound total EP as 
\begin{align}
\EP(p,\protAlpha)&\le\frac{2C(p_{a}'-p_{a})^{2}}{p_{a}'p_{b}}\frac{\ft}{(\ft-1+e^{-\ft})^2}.
\label{eq:appEP67}
\end{align}
Since $\lim_{\ft\to\infty} \ft/(\ft - 1 + e^{-\ft})^2=0$, \cref{eq:appEP67} implies that $\lim_{\ft\to\infty}\EP(p,\protAlpha)\le0$. Since $\EP(p,\protAlpha)\ge0$ by the non-negativity of the EP rate, this proves condition II.

Finally, we show that condition III is satisfied. In particular, we show that  at all $t\in[0,\ft]$,  $\Ralpha$ satisfies LDB, \cref{eq:paramRates}, for some energy function $\Etalpha\in\EE$ (and some arbitrary symmetric activity function $\psi(t)$).  Clearly,  $\Ralpha$ satisfies LDB for all $i,j\not\in\{a,b\}$, which follows by taking $\psi_{ij}(t)=0$. It remains to show that the following conditions hold at all times for some $\Etalpha \in\EE$:
\begin{align*}
 W_{ab}^\ft(t)=f_\ft(t)/C &=\psi_{ab}(t) e^{\beta[E_{a}^{\ft}(t)-E_{b}^{\ft}(t)]/2}\nonumber\\
 W_{ba}^\ft(t)=1-f_\ft(t)/C&=\psi_{ba}(t) e^{\beta[E_{b}^{\ft}(t)-E_{a}^{\ft}(t)]/2}\nonumber\\
 &=\psi_{ab}(t) e^{\beta[E_{b}^{\ft}(t)-E_{a}^{\ft}(t)]/2},
\end{align*}
where we used \cref{eq:trans2} as well as the symmetry $\psi_{ab}(t)=\psi_{ba}(t)$.
Since $\psi(t)$ can be chosen arbitrarily, the above conditions can be restated in terms of the following single equality, 
\begin{equation}
\beta^{-1}\ln\frac{f_{\ft}(t)}{C-f_{\ft}(t)}=E_{b}^{\ft}(t)-E_{a}^{\ft}(t).\label{eq:fg1}
\end{equation}
We now demonstrate that it is possible to choose a sufficiently large $\ft$ such that \cref{eq:fg1} holds at all times $t\in[0,\ft]$.

Recall that $f_{\ft}(t)$ is monotonically decreasing in $t$, so
\begin{align}
\beta^{-1}\ln\frac{f_{\ft}(t)}{C-f_{\ft}(t)} & \le\beta^{-1}\ln\frac{f_{\ft}(0)}{C-f_{\ft}(0)}\nonumber \\
 & =\beta^{-1}\ln\frac{p_{a}}{p_{b}}<g_{ab}(\EE),\label{eq:gf99}
\end{align}
where the last inequality follows from \cref{eq:cg2}. Similarly, 
\begin{align}
\beta^{-1}\ln\frac{f_{\ft}(t)}{C-f_{\ft}(t)}\ge\beta^{-1}\ln\frac{f_{\ft}(\ft)}{C-f_{\ft}(\ft)}.
\label{eq:ggh3}
\end{align}
Since $\lim_{\ft\to \infty} f_{\ft}(t)=p_a$, it follows
that
\begin{align}
\lim_{\ft\to\infty}\beta^{-1}\ln\frac{f_{\ft}(\ft)}{C-f_{\ft}(\ft)}=\beta^{-1}\ln\frac{p_{a}'}{p_{b}'}>-g_{ab}(\EE),
\label{eq:ggh4}
\end{align}
where the last inequality follows from \cref{eq:cg2}. 
Combining \cref{eq:gf99,eq:ggh3,eq:ggh4} implies that for some sufficiently large $\ft$, 
\begin{equation}
-g_{ab}(\EE)<{\beta}^{-1}\ln\frac{f_{\ft}(t)}{C-f_{\ft}(t)}<g_{ab}(\EE)\label{eq:hg3}
\end{equation}
for all $t\in[0,\ft]$. 
Recall from the definition of $g_{ab}(\EE)$ in \cref{eq:egSdef} that
there must be some $E^{(0)},E^{(1)}\in\EE$
such that
\begin{align}
g_{ab}(\EE)\le E_{b}^{(0)}-E_{a}^{(0)},\qquad g_{ab}(\EE)\le E_{a}^{(1)}-E_{b}^{(1)}.
\label{eq:gh3}
\end{align}
Combined with \cref{eq:hg3}, this means that for all $t\in[0,\ft]$,
\begin{equation}
E_{b}^{(1)}-E_{a}^{(1)}<{\beta}^{-1}\ln\frac{f_{\ft}(t)}{C-f_{\ft}(t)}<E_{b}^{(0)}-E_{a}^{(0)}.\label{eq:gf4}
\end{equation}
Finally, since $\EE$ is a path-connected set, there is a continuous curve
of energy functions that connects $E^{(0)}$ and $E^{(1)}$. Given
\cref{eq:gf4} and the intermediate value theorem, 
 there is some $\Etalpha\in\EE$ such that
\cref{eq:fg1} is satisfied for every $t\in[0,\ft]$.

\newcommand{\EspecialA}{E^\star}

\section{Generalization of \cref{lem:transfer} and \cref{thm:a} to non-conservative forces}
\label{app:nonconservative}
\newcommand{\FF}{\mathcal{F}}

In this appendix, we discuss how \cref{lem:transfer,thm:a} can be generalized to the case when there are constraints both on the energy functions and the non-conservative forces that can be applied to a system.

In the presence of non-conservative forces, transitions rates can be  parameterized  via a generalized version of \cref{eq:paramRates},
\begin{align}
\WjiF=\psi_{ji} e^{\beta[\Ex[\ii]-\Ex[\jj] + N_{ji}]/2},
\label{eq:ncarr0}
\end{align}
where $N_{ji}=-N_{ij}$ is an anti-symmetric function that reflects a non-conservative force that biases transitions from state $i$ to state $j$. (As above, $E$  is the energy function and $\psi_{ji}=\psi_{ij}\ge 0$ is a symmetric positive function that controls the overall timescale of transitions between states $\ii$ and $\jj$.)
Now suppose there are constraints on \emph{both} the energy functions and the non-conservative forces that can be applied to a system. For notational convenience, we define an anti-symmetric matrix $F$ such  that $F_{ji}=-F_{ij}$ is the drop of the potential energy  plus the non-conservative bias in going from $i$ to $j$, $F_{ji}:=\Ex[\ii]-\Ex[\jj] + N_{ji}$. 
\cref{eq:ncarr0} can then be written as
\begin{align}
\WjiF=\psi_{ji} e^{\beta[F_{ji}]/2}.
\label{eq:ncarr}
\end{align}
The presence of constraints on the energy functions and non-conservative driving forces can be stated formally as follows:  there is some set $\mathcal{F} \subset \mathbb{R}^{n\times n}$ of anti-symmetric matrices %
such that for any possible driving protocol $\prot=\protSet$, the rate matrix $W(t)$  at all $t\in[0,\ft]$ satisfies \cref{eq:ncarr} for \emph{some} $F\in\FF$.

We now consider how \cref{lem:transfer}, which constructs a protocol that we call a \emph{transfer}, can be generalized to this more general type of constraint. To begin, define the controllable energy gap in terms of $\FF$ (rather than in terms of $\EE$, as in \cref{eq:egSdef}) in the following manner:
\begin{align}
g'_{ij}(\FF) := \min\Big\{ \max_{F\in\FF}\, F_{ij},\; \max_{F\in\FF}\, F_{ji} \Big\}.
\label{eq:NCegSdef}
\end{align} 
Then, in the presence of constraints on both the energy functions and the non-conservative driving forces, \cref{lem:transfer} continues to hold under this new definition. In particular, consider any initial distribution $\ps$ and final distribution $\pf$ which involve transferring $\Delta$ probability from state $a$ to state $b$, as in \cref{eq:dyn0}. Assume that \cref{eq:cg2} holds for this pair of distributions, where $\egS[a][b]$ is replaced by $g'_{ab}(\FF)$ as defined in \cref{eq:NCegSdef}. Then, for any $\epsilon >0$, there is a protocol 
$\prot$ with a time-dependent rate matrix $W(t)$ over $t\in[0,\ft]$ such that: (I) $\ps$ is transformed to $\pf$, (II) $\EP(p,\prot) \le \epsilon$, and (III) $W(t)$ satisfies  \cref{eq:ncarr} for some $F\in\FF$ at all $t$. 
The proof of statements (I) and (II) is exactly the same as appears in \cref{app:transfer}. The proof of statement (III) is also the same as appears in  \cref{app:transfer}, as long as the following  replacements are made:
\begin{enumerate}
\item  $\egS$ should be replaced by $g'_{ij}(\FF)$,
\item Statements like ``$E(t)\in \EE$'' should be replaced by statements like ``$W(t)$ obeys \cref{eq:ncarr} for some $F\in\FF$'',
\item Expressions like $E_b - E_a$ in \cref{eq:fg1,eq:gh3,eq:gf4} should be replaced by corresponding versions with $F_{ba}$. 
\end{enumerate}

Finally, note that  \cref{thm:2} is proved (informally in the main text) by an explicit construction that shows how the initial distribution $\ps$ can be transformed to a final distribution $\pf$ via an appropriate sequence of transfers. The same construction also works for transfers defined under constraints on both energy functions and non-conservative driving forces (as discussed above), as long as the capacity term $C$ which appears in \cref{eq:cond1,eq:mst2,eq:maxCapDef} is defined in terms of controllable energy gaps $g'_{ij}(\FF)$ from \cref{eq:NCegSdef}, rather than $\egS$.

\section{Derivation of \cref{eq:eucCap}}
\label{app:eucCap}

Suppose that $\EE$ contains a one-dimensional curve of energy functions that connects $\Enot$ and $-\Enot$. Consider some pair of states $\ii,\jj$, and first assume that $\EnotX[\ii]\ge \EnotX[\jj]$. Then, using the definition of $\egS$ in \cref{eq:egSdef},
\begin{align*}
\egS &\ge \min\Big\{ \EnotX - \EnotX[\jj],(-\EnotX[\jj]) - (-\EnotX[\ii])\Big\}=\EnotX - \EnotX[\jj].
\end{align*}
Conversely, if $\EnotX[\ii]\le \EnotX[\jj]$, then $\egS \ge \EnotX[\jj]-\EnotX[\ii]$. Combining these results implies
\begin{align}
\label{eq:absn}
\egS \ge  |\EnotX[\ii] - \EnotX[\jj]|.
\end{align}
Next,  for any state $\ii$, let $m(\ii)$ indicate the state with the largest energy difference  from $\ii$ under $\Enot$,
\begin{align}
m(\ii) \in \argmax_{k}  |\EnotX[\ii] - \EnotX[k]|.
\label{eq:miidef}
\end{align}
Given a pair of states $\ii,\jj$, we consider two possibilities. Under the first possibility, $\EnotX[m(\ii)]=\EnotX[m(\jj)]$ and  the path $\vec{x} = (\ii \to m(\ii) \to \jj)$ obeys
\begin{align}
\pathcap(\vec{x})&=\min \{ \egS[\ii][m(\ii)], \egS[m(\ii)][\jj]\} \nonumber \\
&\ge \min \{|\EnotX[\ii] - \EnotX[m(\ii)]|,|\EnotX[m(\ii)] - \EnotX[\jj] | \} \nonumber \\
&= \min \{|\EnotX[\ii] - \EnotX[m(\ii)]|,|\EnotX[m(\jj)]-\EnotX[\jj] | \} ,\label{eq:appG7}
\end{align}
where in the second line we used \cref{eq:absn}. 
Under the second possibility,  $\EnotX[m(\ii)]\ne \EnotX[m(\jj)]$ and capacity of the path  $\vec{x} = (\ii \to m(\ii) \to m(\jj) \to \jj)$ obeys
\begin{multline}
\pathcap(\vec{x})=\min \{  \egS[\ii][m(\ii)],\egS[m(\ii)][m(\jj)],\egS[m(\jj)][\jj]\} \ge \\
  \min \{ |\EnotX[\ii] - \EnotX[m(\ii)]|,|\EnotX[m(\ii)] - \EnotX[m(\jj)]|,|\EnotX[m(\jj)] - \EnotX[\jj]|\}.
\label{eq:gh5}
\end{multline}
Observe that for any $\ii$, the state $m(\ii)$ defined in \cref{eq:miidef} either obeys  $m(\ii) \in \argmin_\jj \EnotX[\jj]$ or $m(\ii) \in \argmax_\jj \EnotX[\jj]$. Since we assumed that  $\EnotX[m(\ii)]\ne \EnotX[m(\jj)]$, it must be that the set $\{m(\ii),m(\jj)\}$ includes both lowest and highest energy states. This in turn implies that $|\EnotX[m(\ii)] - \EnotX[m(\jj)]|$ is  larger than both $ |\EnotX[m(\ii)] - \EnotX[\ii]|$ and $ |\EnotX[m(\jj)] - \EnotX[\jj]|$, allowing  us to rewrite \cref{eq:gh5} as
\begin{align}
\pathcap(\vec{x})&\ge \min \{ |\EnotX[\ii] - \EnotX[m(\ii)]|,|\EnotX[m(\jj)] - \EnotX[\jj]|\}, \label{eq:appG8}
\end{align}
which is the same as \cref{eq:appG7}.

To summarize, above we showed that for any pair of states $\ii,\jj$, there exists a path $\vec{x}$ from $\ii$ to $\jj$ such that
\begin{align}
\pathcap(\vec{x}) & \ge \min \{ |\EnotX[\ii] - \EnotX[m(\ii)]|,|\EnotX[m(\jj)] - \EnotX[\jj]|\}\nonumber\\
&=\min \{\max_k |\EnotX[\ii] - \EnotX[k]|,\max_k |\EnotX[\jj] - \EnotX[k]| \}\nonumber\\
&\ge \min_l \max_k |\EnotX[l] - \EnotX[k]|,
 \label{eq:ggd9}
\end{align}
where we used \cref{eq:appG7,eq:appG8} and the definition in \cref{eq:miidef}. The first inequality in \cref{eq:eucCap} then follows immediately from \cref{eq:ggd9} and the definition of $\maxCap$ in \cref{eq:maxCapDef}:
\begin{align}
\maxCap &= \min_{\ii \ne \jj} \max_{\vec{x} \in \mathcal{P}(\ii,\jj)} \pathcap(\vec{x}) \label{eq:maxCapApp0} \\
&\ge \min_{\ii} \max_{\jj} |\EnotX[\ii] - \EnotX[\jj]|.
\label{eq:maxCapApp}
\end{align}
To derive the second inequality in \cref{eq:eucCap}, we consider two cases. First,  for any $i$ such that $\EnotX[\ii] \le \frac{\max_j \EnotX[\jj] + \min_j \EnotX[\jj]}{2}$, 
\begin{align*}
\max_{\jj} |\EnotX[\ii] - \EnotX[\jj]| = \max_j \EnotX[\jj] - \EnotX[\ii]\ge \frac{\max_j \EnotX[\jj] - \min_j \EnotX[\jj]}{2}.
\end{align*}
Second,  for any $i$ such that $\EnotX[\ii] > \frac{\max_j \EnotX[\jj] + \min_j \EnotX[\jj]}{2}$,
\begin{align*}
\max_{\jj} |\EnotX[\ii] - \EnotX[\jj]| = \EnotX[\ii]-\min_j \EnotX[\jj] \ge \frac{\max_j \EnotX[\jj] - \min_j \EnotX[\jj]}{2}.
\end{align*}

We finish by showing that equality is achieved in \cref{eq:maxCapApp}
if the set of energy functions is a one-dimensional line segment, \[
\EE = \{ \lambda \Enot  : \lambda \in [-1,1]\}.\]
In that case, it is easy to verify from 
 the definition of $\egS$ that %
  $\egS =  |\EnotX[\ii] - \EnotX[\jj]|$, so equality is achieved in \cref{eq:absn}. Next, note that for any path $\vec{x}$ that starts or ends on state $\ii$, it must be that $$\pathcap(\vec{x})\le \max_{\jj}\egS=\max_{\jj} |\EnotX[\ii] - \EnotX[\jj]|.$$ Plugging into the definition of $\maxCap$ (see \cref{eq:maxCapApp0}) and simplifying gives $$\maxCap \le \min_{\ii} \max_{\jj} |\EnotX[\ii] - \EnotX[\jj]|,$$ which implies equality in \cref{eq:maxCapApp}.

\section{Derivation of \cref{eq:ineq34}}
\label{app:eprateEtaBound}

Assume that  \cref{eq:optA} holds for all $p$ and $\R \in \RR$.  Then, it must hold for the distribution and rate matrix at all times $t\in[0,\ft]$, $p(t)$ and $\R(t)$.  The  right side of  \cref{eq:optA} can be written in terms  of the following time derivative,
\begin{multline}
-\mult\sum_{\ii,\jj} \pxt \WjiF(t) \ln\frac{\pxt[\jj]}{[\klprojT]_\jj}\\
=-\eta\ddt \DKL(\pt\Vert q)\big\vert_{q=\klprojT}.
\label{eq:z0}
\end{multline}
The total
derivative rule from calculus then gives 
\begin{align}
&-\ddt \DKL(\pt\Vert q)\vert_{q=\klprojT}=\nonumber\\
&\;\;-\ddt D(\pt\Vert\klprojT)+\ddt \DKL(q\Vert\klprojT)\vert_{q=\pt}.\label{eq:z1}
\end{align}
The second term above can be bounded as 
\begin{align*}
&\ddt D(q\Vert\klprojT)\vert_{q=\pt}=\\
&\;\;\;\lim_{s\to0}\frac{1}{s}[\DKL(\pt\Vert\klproj[{e^{s\R(t)}\pt}]-\DKL(\pt\Vert\klprojT)]\ge0,
\end{align*}
where the inequality follows from the definition of the KL projection in \cref{eq:klproj}.
Combining with \cref{eq:z1} gives 
\[
-\ddt \DKL(\pt\Vert q)\vert_{q=\klprojT}\ge-\ddt \DKL(\pt\Vert\klprojT),
\]
which can be combined with \cref{eq:optA,eq:z0} to give \cref{eq:ineq34}. 

\section{Finding $\eta$ in \cref{thm:2} with convex optimization\label{app:convexopt}}

In this appendix, we show how to identify the optimal value of $\mult$
for the EP bound in \cref{thm:2} using convex optimization techniques. 
 We use $\allDistSet$ to indicate the set of all probability distribution over the state space of the system.  In addition, we use the general definition of the KL projection to some set of distributions $\distSet$, 
\begin{align}
\klproj:=\argmin_{q \in \distSet }\DKL(p\Vert q ).\label{eq:klproj2}
\end{align}
Here we assume that $\distSet$ is a finite set.

For notational convenience, define the following function: %
\begin{align}
&f(p,q,\R,\mult)=\label{eq:appcp2} \\
&(1-\eta)\Big[-\sum_{\ii,\jj}\px \WjiF\ln {\px[\jj]} \Big]+ \sum_{\ii,\jj}\px \WjiF\Big[\ln\frac{\WjiF}{\WijF}-\eta\ln q_\jj \Big],
\nonumber
\end{align}
where $p\in\allDistSet $ and $q\in\allDistSet $ are probability distributions, $\R$ is a 
rate matrix, and $\mult\in[0,1]$
is some scalar. 
Note that the first bracketed term on the right hand side of \cref{eq:appcp2} is the rate of increase of the Shannon entropy  of distribution $p$ under rate matrix $\R$. This term is convex in $p$, since it differs from the EP rate by a linear function of $p$ (see \cref{eq:appcp0} below), and the EP rate is convex in $p$~\cite{maes2013minimum}. 
Thus, for a fixed $\R$, $E$,
and $\mult$, %
$f$ is a weighted sum of a convex function 
of $p$ and a linear function of $p$, meaning that it is a convex function of $p$. 

For any distribution $p$ and rate matrix $\R$, rewrite the left
hand side of \cref{eq:optA} (the EP rate incurred by distribution $p$ under rate matrix $\R$) as
\begin{align}
-\sum_{\ii,\jj}\px \WjiF\ln {\px[\jj]}+\sum_{\ii,\jj}\px\WjiF\ln\frac{\WjiF}{\WijF},\label{eq:appcp0}
\end{align}
where we used \[
\sum_{\ii,\jj}\px \WjiF\ln {\px}=\sum_{\ii}\px \Big(\sum_\jj \WjiF \Big)\ln {\px}=0.
\]
Next, rewrite the right hand side of \cref{eq:optA}
as
\begin{align}
-\eta\sum_{\ii,\jj}\px \WjiF\ln \px[\jj]+\eta\sum_{\ii,\jj}\px\WjiF\ln[\klproj]_\jj .
\label{eq:appcp3}
\end{align}
We can now use \cref{eq:appcp0,eq:appcp2,eq:appcp3} to restate \cref{eq:optA}
in the following way: \cref{eq:optA} holds for a given $\mult\in[0,1]$
if for all distributions $p$, $\R\in\RR$, and $q \in\distSet$ such that $\klproj=q$,
\begin{equation}
f(p,q, \R,\mult)\ge0.\label{eq:optProb1}
\end{equation}
Observe that $\klproj=q$ is equivalent to the condition 
\begin{equation}
\DKL(p\Vert q)\le \DKL(p\Vert q')\quad\forall q'\in\distSet,\label{eq:nm9}
\end{equation}
which, with some simple rearranging, is equivalent to the following
set of linear constraints on $p$,
\[
\forall q'\in\distSet:\sum_{\ii}\px\ln\frac{q_\ii}{q'_\ii}\ge0.
\]

To summarize, \cref{eq:optA} holds for a given $\mult$
when the following inequality is satisfied for each $\R\in\RR$ and $q\in\distSet$:
\begin{align}
\begin{aligned}
0 \le & \;\min_{p\in\allDistSet} f(p,q,\R,\mult) \\
&\text{\;\;\;s.t.\;} \forall q'\in\distSet: \sum_{\ii}\px\ln\frac{q_\ii}{q'_\ii}\ge 0.
\end{aligned}
\label{eq:optapp}
\end{align}
For each $\R$ and $q$, 
\cref{eq:optapp} can be verified for a given $\mult$ by solving a convex minimization problem 
subject to linear constraints, which can be done efficiently using
standard numerical techniques. 
For a finite set of rate matrices $\RR$ and distributions $\distSet$, one can solve $\vert\RR\vert \times \vert\distSet\vert$ such problems to verify whether \cref{eq:optA} holds for a given $\mult$.
Finally, to find the largest such $\mult\in[0,1]$
(thereby making \cref{eq:optA} as tight as
possible), one can use the bisection method on the interval $\mult\in[0,1]$.
Code for doing this optimization is available at \url{https://github.com/artemyk/epbound}.

\section{Generalization of \cref{thm:2} to multiple reservoirs and/or non-conservative forces}
\label{app:thm2multiplebaths}

Our derivation \cref{thm:2} makes no assumptions about the structure of the rate matrices in $\RR$, including whether or not local detailed balance is obeyed. For this reason, the results derived in \cref{sec:second} can be  generalized to consider systems coupled to multiple reservoirs and/or subject to non-conservative forces. 
In this more general situation, $\klproj[\cdot]$ should be defined as the projection to the set of non-equilibrium stationary distributions of the rate matrices in $\RR$ (or alternatively, as mentioned in \cref{sec:second}, to any other set of distributions). \cref{thm:2} then applies for any $\eta\in[0,1]$ which satisfies \cref{eq:optA} for this $\klproj[\cdot]$.

In addition, the bound in \cref{thm:2} can be strengthened for systems coupled to multiple reservoirs, as long as each rate matrix  in $\R\in\RR$ can be decomposed into separate contributions from each reservoir $r$ as $\R=\sum_r \R^{(r)}$~\cite{esposito2010three}. In that case,  the EP rate (which appears on the left hand side of \cref{eq:optA}) can defined as~\cite{esposito2010three} 
\begin{align} 
\EPr(p,W) &= \frac{1}{2}\sum_{\ii, \jj,r}{ (\px\WjiF^{(r)}- \px[\jj]\WijF^{(r)})}  \ln \frac{\px\WjiF^{(r)}}{\px[\jj]\WijF^{(r)}}.
\label{eq:epratemult0}
\end{align}
(Compare to \cref{eq:eprate0} in the main text, which applies in the presence of a single reservoir.) 
The expression in \cref{eq:epratemult0} is an upper bound on the expression in \cref{eq:eprate0}~\cite{esposito2010three}, so the largest $\mult$ which satisfies the implicit inequality \cref{eq:optA} for $\EPr$ as defined in \cref{eq:epratemult0} will be no smaller than (and possibly larger than) the largest $\mult$ which satisfies that inequality for $\EPr$ as defined in \cref{eq:eprate0}.

\clearpage
\end{document}